\documentclass[10pt]{article}

\usepackage{amsfonts,amsmath,amssymb,amsthm}
\usepackage[pdftex]{graphicx}
\usepackage[hmargin=1in,vmargin=1in]{geometry}
\usepackage{natbib,setspace,enumerate,mathtools}
\usepackage{multirow}
\usepackage{booktabs}
\allowdisplaybreaks
\usepackage[toc,title,titletoc,header]{appendix}
\usepackage{adjustbox}
\usepackage[colorlinks,citecolor=blue]{hyperref}
\usepackage{natbib} 
\usepackage{etoolbox} 
\usepackage{threeparttable}
\usepackage{tikz}  \usetikzlibrary{shapes.geometric}
\usepackage{marvosym}

\def\qed{\rule{2mm}{2mm}}

\theoremstyle{definition}

\AtEndEnvironment{remark}{~\qed}
\AtEndEnvironment{example}{~\qed}

\DeclareMathOperator{\var}{Var}
\DeclareMathOperator{\cov}{Cov}

\newcommand{\mycomment}[1]{}

\begin{document}

\author{
Yuehao Bai \\
Department of Economics\\
University of Southern California \\
\url{yuehao.bai@usc.edu}
\and
Azeem M.\ Shaikh\\
Department of Economics\\
University of Chicago \\
\url{amshaikh@uchicago.edu}
\and
Max Tabord-Meehan\\
Department of Economics\\
University of Chicago \\
\url{maxtm@uchicago.edu}
}

\bigskip

\title{A Primer on the Analysis of Randomized Experiments and a Survey of some Recent Advances \thanks{We thank the co-editor and an anonymous referee for comments and suggestions that have improved the manuscript. We would also like to thank Gustavo Bobonis for helpful discussions. The third author acknowledges support from NSF grant SES-2149408.}}

\maketitle

\section{Introduction}

The past two decades have witnessed a surge of new research in the analysis of randomized experiments.  The emergence of this literature may seem surprising given the widespread use and long history of experiments as the ``gold standard'' in program evaluation, but this body of work has revealed many subtle aspects of randomized experiments that may have been previously unappreciated. This article provides an overview of some of these topics, primarily focused on stratification, regression adjustment, and cluster randomization, although we also provide short discussions on a broad range of other topics at the end of the article. We also provide a companion {\tt R} package {\tt sreg}, which is designed to facilitate inference in stratified (as well as potentially clustered) randomized experiments: see \cite{trifonov2025sreg}.

The majority of our discussion is presented within the context of a framework in which units are sampled from a suitable ``super-population''; i.e., we assume that the units are i.i.d.\ draws from some probability distribution.  Importantly, these results differ from results that are derived within a complementary framework in which units are sampled from a fixed, finite population (without replacement).  The special case in which all the units from the finite population are sampled is sometimes referred to as \emph{design-based} inference.  After defining some common notation in Section \ref{sec:setup}, we begin our review in Section \ref{sec:fin/super} by providing a comparison of these two sampling frameworks.  Specifically, we introduce both frameworks in the context of an analysis of the difference-in-means estimator for the average treatment effect in a completely randomized experiment.  There, we argue that the super-population framework approximates a finite-population framework in which a negligible fraction of the finite population is sampled; we argue further, however, that results derived in a super-population framework may provide useful methods for inference in a finite population even outside of this limiting case.


We then turn our attention in Section \ref{sec:sbr} to the impact of stratification on subsequent inferences drawn from randomized experiments.  More specifically, instead of assigning treatment status according to complete randomization, it is common to stratify first according to some baseline covariates and then assign treatment status within each stratum so as to ensure that the treatment group and control group in the experiment are ``balanced'' according to these covariates.  Examples of such schemes include stratified block randomization and matched pair designs, both of which are commonly used within economics and the sciences more generally:  see \cite{rosenberger2015randomization} for a textbook treatment focused on clinical trials and \cite{bruhn2009pursuit} for a review focused on development economics. We first illustrate in Section \ref{sec:strat_benefits} how stratification can improve the precision of the usual difference-in-means estimator of the average treatment effect by reducing what we refer to as the ex-post (as opposed to ex-ante) bias of the estimator.  Here, the ex-post bias of the estimator is used to describe its behavior conditional on treatment status rather than unconditionally.  Section \ref{sec:strat_inference} develops the implications of this increase in precision for inferences about the average treatment effect more formally for the special case of stratified block randomization with finitely many ``large" strata;  Section \ref{sec:small_strata} describes how this analysis changes when strata are ``small," for instance in the case of matched pair designs.

In Section \ref{sec:cov-adjust} we consider the use of baseline covariates (including ones that are possibly not used in the assignment of treatment status) to further improve the precision of estimators of the average treatment effect through regression adjustment. We explain how na\"ive regression adjustment may increase rather than decrease the precision of estimators of the average treatment effect, and how a more careful use of the covariates can ensure an improvement in (asymptotic) precision. Our discussion includes linear adjustments (Section \ref{sec:adjust-example}) as well as more general forms of adjustment (Section \ref{sec:adjust-general}). 


Finally, Section \ref{sec:clust} extends our discussion to cluster randomized experiments, i.e., randomized experiments in which the unit of randomization is a cluster.  Such designs are increasingly common in economics.  Indeed, \cite{muralidharan2017experimentation} find in a survey of leading economics journals between 2001 and 2016 that more than 75 percent of randomized experiments were cluster randomized.  In Section \ref{sec:clust_params} we draw a distinction between different ways in which one may define the average effect of the treatment in such settings.  In particular, we discuss two possible parameters of interest, which we call the equally-weighted average treatment effect and the size-weighted average treatment effect, that differ in how the average effect within a cluster is aggregated across individuals. In Section \ref{sec:clust_analysis} we discuss inference in cluster randomized experiments.

Owing to space constraints, our discussion is, of course, necessarily incomplete. We therefore provide, at the end of each section, a guide to some of the related literature on similar topics.  For completeness, in Section \ref{sec:other} we also briefly discuss a number of topics that were regrettably omitted from the main discussion: this includes the analysis of treatment effect heterogeneity, re-randomization, multiple testing, imperfect compliance/attrition, experiments with interference, randomization inference, policy learning, and response-adaptive designs. \cite{athey2017econometrics} also provide a review of some of these topics (primarily from the design based perspective).

Before proceeding, we emphasize that our survey is limited to the \emph{analysis} of randomized experiments, and as a result we will only briefly comment on some aspects of experimental \emph{design} as well as practical issues surrounding \emph{implementation}, in passing; classical textbook treatments on experimental design are provided in \cite{cox2000theory},  \cite{pukelsheim2006optimal}, \cite{atkinson2007optimum}, and \cite{wu2011experiments}.  Other important contributions to the theory of experimental design (including theoretical justifications for \emph{why} an experimenter may want to randomize) are provided in \cite{savage1951theory}, \cite{blackwell1954theory}, \cite{kiefer1959optimum}, \cite{li1983minimaxity},  \cite{kallus2018optimal,kallus2021optimality}, Section 5.10 in \cite{lehmann2022testing}, and \cite{bai2023why}. Important references discussing other practical issues (particularly for field experiments conducted in economics) include \cite{duflo2007using}, \cite{glennerster2013running}, \cite{karlan2017failing}, and  \cite{list2023voltage}.


\section{Setup and Notation: The Potential Outcomes Framework}\label{sec:setup}
In this section, we present some notation which will be common to the majority of the article. Each individual $i$ in the experiment is assigned a binary treatment $D_i \in \{0, 1\}$ (we focus on settings with binary treatments, but provide references on related extensions to multiple treatments throughout the article). Let $Y_i(1)$ denote the potential (or counterfactual) outcome for individual $i$ if they are treated, and $Y_i(0)$ denote the potential outcome if they are untreated. Note that we never observe $Y_i(1)$ and $Y_i(0)$ simultaneously for the same individual, but rather we observe the outcome $Y_i$ given by
\[ Y_i = \begin{cases}
    Y_i(1) & \text{ if } D_i = 1 \\
    Y_i(0) & \text{ if } D_i = 0~.
\end{cases} \]
We summarize the previous relationship succinctly by
\[ Y_i = Y_i(1) D_i + Y_i(0) (1 - D_i)~. \]
For each individual, we may also observe a vector of baseline covariates denoted by $X_i$. The experimental sample is thus given by $\{(Y_i, D_i, X_i): 1 \le i \le n\}$. For any variable indexed by $i$, for example $D_i$, we denote by $D^{(n)}$ the vector $(D_1, D_2, \ldots, D_n)$. Note that we will always model the experimental assignments $D^{(n)}$ as random; on the other hand, as we will explain below, depending on the sampling framework employed in the analysis, $\{(Y_i(1), Y_i(0), X_i): 1 \le i \le n\}$ may either be modelled as random vectors or fixed quantities.

Much of our discussion will center on the properties of the standard difference-in-means estimator: let $n_1 = \sum_{1 \le i \le n} D_i$  denote the number of treated units in the sample and $n_0 = n - n_1$ denote the number of control units. The difference-in-means estimator is then given by 
\[\hat{\Delta}_n = \frac{1}{n_1}\sum_{1 \le i \le n}Y_iD_i - \frac{1}{n_0}\sum_{1 \le i \le n}Y_i(1-D_i)~.\]
 Note that this estimator can equivalently be described as the estimator of the coefficient on $D_i$ when estimating the following linear regression by least squares:
\begin{equation}\label{eq:OLS}
\texttt{regress } Y_{i} \texttt{ on } \text{constant} +  D_i~.
\end{equation}

We will illustrate many of the basic concepts via the analysis of a \emph{completely randomized} experiment. In a completely randomized experiment, the treatment assignment $D^{(n)}$ is implemented such that, for some fixed fraction of units, say $\pi$, exactly $n_1 = \lfloor\pi n \rfloor$ units are assigned to treatment and $n_0 = n - n_1$ units are assigned to control, with all such possible assignments being equally likely. Formally, let $(d_1, \ldots, d_n) \in \{0, 1\}^n$, then $D^{(n)}$ is independent of $\{(Y_i(1), Y_i(0), X_i): 1 \le i \le n\}$ with distribution given by
\[
P\{D^{(n)} = (d_1, \ldots, d_n)\} =  
\begin{cases} 
\binom{n}{n_1}^{-1} & \text{if } \sum_{1 \le i \le n} d_i = n_1~, \\
0 & \text{otherwise}~.
\end{cases}
\]
We emphasize that an important feature of our discussion here and throughout the rest of the article is that we do not assume that the components of $D^{(n)}$ are independently distributed; this will play a crucial role in our subsequent analyses when studying, for example, stratified randomization in Section \ref{sec:sbr}.

\section{What is Random? Finite versus Super-population Analyses of a Completely Randomized Experiment}\label{sec:fin/super}

In this section, we introduce the two main paradigms for the analysis of randomized experiments: the finite-population and super-population approaches to inference. In a finite-population analysis, we begin with a collection $\{(y_j(1), y_j(0), x_j): 1 \le j \le N\}$ of \emph{fixed} quantities that constitute the entire population of interest. A sample of size $n \le N$, given by $\{(Y_i(1), Y_i(0), X_i): 1 \le i \le n\}$, is then drawn \emph{without} replacement from the population, and the experiment is performed on these $n$ individuals. The most common case considered in the literature is when $n = N$, so that the experiment is performed on the entire population. We will refer to this case as \emph{design-based} inference, since the only source of uncertainty arises from the randomness in $D^{(n)}$. This perspective is often considered attractive in settings where it is difficult to conceptualize an appropriate sampling frame; \cite{reichardt1999justifying} provide further discussion.

In contrast, in a super-population analysis, the sample $\{(Y_i(1), Y_i(0), X_i): 1 \le i \le n\}$ is modeled as being i.i.d.\ according to some probability distribution. Informally, we may view the distribution from which the potential outcomes are drawn as summarizing an essentially infinite ``super-population.'' 

To compare and contrast the super- and finite population paradigms, we present an analysis of the difference-in-means estimator $\hat{\Delta}_n$ for the average treatment effect under both approaches, following the original work of \cite{neyman1923application}, in a completely randomized experiment (as defined in Section \ref{sec:setup}). Our primary takeaway will be that, in a sense to be made formal below, we can view the super-population paradigm as an approximation to the finite-population paradigm in a regime where the sample size $n$ is a vanishing fraction of the total population size $N$. Consequently, we will show that in the super-population framework it is possible to construct \emph{consistent} variance estimators of $\var[\hat{\Delta}_n]$, whereas this will often be impossible in the finite-population framework. Instead, in the finite population framework we will explain how to construct \emph{conservative} variance estimators of $\var[\hat{\Delta}_n]$. Moreover, we will argue that consistent variance estimators derived via a super-population analysis are often reasonable conservative variance estimators when viewed through the lens of a finite-population analysis.


\subsection{Finite population analysis of $\hat{\Delta}_n$}\label{sec:finpop_intro}

First we consider a finite-population analysis. Recall that in this case we begin with $\{(y_j(1), y_j(0), x_j): 1 \le j \le N\}$ which are \emph{fixed}, non-random quantities that describe the outcomes (and covariates) of the \emph{entire} population of $N$ units. Accordingly, in this case, the average treatment effect is defined as
\[\Delta^{\rm fp}_N = \frac{1}{N}\sum_{1 \le j \le N}\left(y_j(1) - y_j(0)\right)~.\]
We emphasize that in this framework the parameter of interest $\Delta^{\rm fp}_N$ is determined entirely by the values of the potential outcomes from the $N$ units in the population. 


To analyze $\hat{\Delta}_n$ from a finite population perspective, it is often useful to re-frame the problem as a problem of survey sampling from a finite population; re-framing the problem in this way allows us to employ classical results from survey sampling \citep[see, for instance,][]{cochran1977sampling, lehmann2022testing}. In Appendix \ref{sec:fin_bias_var} we illustrate how this re-framing can be useful in deriving some ``finite-$N$" properties of $\hat{\Delta}_n$ (i.e., properties that hold for every finite population size $N$) in a completely randomized experiment. In particular, there we show 
\[E[\hat{\Delta}_n] = \Delta^{\rm fp}_N~,\]
i.e., that $\hat{\Delta}_n$ is an unbiased estimator for $\Delta^{\rm fp}_N$, and 
\begin{equation}\label{eq:fp_var}
\var[\hat{\Delta}_n] = \frac{S^2_1}{n_1} + \frac{S^2_0}{n_0} - \frac{S^2_\Delta}{N}~,
\end{equation}
where
\[S^2_d = \frac{1}{N-1}\sum_{1 \le j \le N}\left(y_j(d) -\bar{y}_N(d)\right)^2\]
\[S^2_\Delta = \frac{1}{N-1}\sum_{1 \le j \le N}\left(y_j(1) - y_j(0) - \Delta^{\rm fp}_N\right)^2~,\]
and $\bar y_N(d) = \frac{1}{N}\sum_{1 \le j \le N} y_N(d)$.

This expression for the variance of $\hat{\Delta}_n$ is fundamental to understanding why inference on $\Delta^{\rm fp}_N$ is generally conservative in the finite population paradigm: note that $S^2_d$, $d \in \{0, 1\}$ are simply the variances for the potential outcomes in the finite population, and we explain below how to construct estimators of these quantities.  To rule out degenerate situations, we henceforth assume that at least one of  $S^2_d$, $d \in \{0, 1\}$ are nonzero. The quantity $S^2_\Delta$, however, is the  variance of the unit-level treatment effects $(y_j(1) - y_j(0): 1 \le j \le N)$ in the finite population, and these are by definition \emph{never} observed for a given unit. In settings where the experiment size $n$ is a non-trivial fraction of the population size $N$ (for instance, in design-based inference when $n = N$), this feature of the variance introduces an unavoidable roadblock for estimating $\var[\hat{\Delta}_n]$ consistently,\footnote{Here and throughout the rest of Section \ref{sec:finpop_intro}, consistency should be understood as saying that the \emph{ratio} of the estimator to the variance converges to one in probability.} and thus estimators of $\var[\hat{\Delta}_n]$ will necessarily be conservative unless treatment effects are constant (i.e., that $y_j(1) - y_j(0) = \Delta^{\rm fp}_N$ for every $1 \le j \le N$). In cases where $n$ is a vanishing fraction of $N$, however, we see that $S^2_{\Delta}/N$ becomes a negligible component of $\var[\hat{\Delta}_n]$ as $N$ gets large, in the sense that in order to estimate $\var[\hat{\Delta}_n]$ consistently it is sufficient to estimate only the sum of its first two terms consistently. As we show in Section \ref{sec:superpop_intro}, this feature of $\var[\hat{\Delta}_n]$ when $n$ is a vanishing fraction of $N$ will exactly mirror our findings in the super-population analysis.

Before discussing estimation of $\var[\hat{\Delta}_n]$ and related methods of inference for $\Delta^{\rm fp}_N$, we document some necessary ``large-$N$" properties of $\hat{\Delta}_n$ in the finite population paradigm. In the finite population framework we conceptualize our asymptotic approximations by imagining a sequence of increasingly larger populations on which we perform our experiment. As a consequence, our large-$N$ results will require some discipline on how this sequence of ever larger populations evolve as a function of $N$. Under appropriate assumptions on the sequence of populations, it can be shown that \citep[see, e.g.,][Theorem 12.2.5]{lehmann2022testing}
\[\frac{\hat{\Delta}_n - \Delta^{\rm fp}_N}{\sqrt{\var[\hat{\Delta}_n]}} \xrightarrow{d} N(0, 1)~,\]
as $n \rightarrow \infty$ and $N \rightarrow \infty$. Using this result, asymptotic inference on $\Delta^{\rm fp}_N$ is straightforward once we have an estimator of $\var[\hat{\Delta}_n]$. Motivated by our decomposition in \eqref{eq:fp_var}, a conservative variance estimator can be constructed by simply estimating the following upper bound on $\var[\hat{\Delta}_n]$:
\[V_n^{\rm obs} = \frac{S^2_1}{n_1} +  \frac{S^2_0}{n_0}~,\] 
which implicitly sets $S^2_{\Delta}$ to its lowest possible value of zero. From this expression we see that a consistent estimator of $V_n^{\rm obs}$ can be obtained by replacing $S^2_1$ and $S^2_0$ by their natural estimators. Equivalently, viewing $\hat{\Delta}_n$ as the estimator of the coefficient on $D_i$ obtained by the regression in \eqref{eq:OLS}, it can be shown that a consistent estimator of $V_n^{\rm obs}$ can simply be obtained from the resulting heteroskedasticity-robust variance estimator \citep[see, e.g.,][Chapter 8]{angrist2009mostly}. An asymptotically valid $95\%$-confidence interval for $\Delta_N^{\rm fp}$ can therefore be constructed as
\begin{equation}\label{eq:Cn}
C_n = \left[\hat{\Delta}_n - 1.96\cdot \mathrm{SE}(\hat{\Delta}_n), \hat{\Delta}_n + 1.96\cdot \mathrm{SE}(\hat{\Delta}_n)\right]~,
\end{equation}
where $\mathrm{SE}(\hat{\Delta}_n)$ is the robust standard error of the coefficient on $D_i$ obtained from the regression in \eqref{eq:OLS}. Although $C_n$ is a valid confidence interval, we emphasize that it is conservative in the sense that 
\[P\{\Delta^{\rm fp}_N \in C_n\} \rightarrow p \ge 0.95~,\]
as $n, N \rightarrow \infty$, with equality only if $n/N \rightarrow 0$ or $S^2_\Delta = 0$.

We conclude this section by considering the following natural follow-up question: could we develop more precise methods of inference by constructing less conservative variance estimators of $\var[\hat{\Delta}_n]$? For the example we just presented, this would amount to considering tighter lower bounds for $S^2_\Delta$ which are themselves consistently estimable. For instance, it follows from the Cauchy-Schwarz inequality that 
\[ \sum_{1 \leq j \leq N} (y_j(1) - \bar y_N(1)) (y_j(0) - \bar y_N(0)) \leq \Bigg ( \sum_{1 \leq j \leq N} (y_j(1) - \bar y_N(1))^2 \Bigg )^{1/2} \Bigg ( \sum_{1 \leq j \leq N} (y_j(0) - \bar y_N(0))^2 \Bigg )^{1/2}~, \]
 which we can use to immediately verify the lower bound $S^2_\Delta \ge (S_1 - S_0)^2 \ge 0$. We thus obtain the following improved upper bound on $\var[\hat{\Delta}_n]$:
\begin{equation} \label{eq:improved}
\frac{S^2_1}{n_1} +  \frac{S^2_0}{n_0} - \frac{(S_1 - S_0)^2}{N}~,     
\end{equation}
which can be used to construct a less conservative estimator of $\var[\hat{\Delta}_n]$. In fact, it is possible to achieve even tighter upper bounds by further exploiting the structure of $S^2_\Delta$: see \cite{aronow2014sharp} for details. We emphasize, however, that although we will argue in Section \ref{sec:superpop_intro} that inferences based on the heteroskedasticity-robust standard error $\mathrm{SE}(\hat{\Delta}_n)$ as in \eqref{eq:Cn} will be valid under either the super-population or finite population paradigms, design-based standard errors (that is, standard errors which are valid in the finite population setting when $n = N$) constructed using improved upper bounds like \eqref{eq:improved} will generally be too small, and thus \emph{invalid} when viewed from a super-population perspective.

\subsection{Super-population analysis of $\hat{\Delta}_n$}\label{sec:superpop_intro}
Next, we repeat the above analysis within the super-population paradigm. Recall that in this case the sample $\{(Y_i(1), Y_i(0), X_i): 1 \le i \le n\}$ is modeled as being i.i.d.\ according to some probability distribution. Accordingly, in this case the average treatment effect is defined as 
\[\Delta = E[Y_i(1) - Y_i(0)]~,\]
where the expectation is with respect to the distribution of the data.

We begin by documenting some finite-sample properties of the estimator $\hat\Delta_n$ in a completely randomized experiment under the super-population framework, analogous to the properties derived in Section \ref{sec:finpop_intro}.  Using familiar properties of conditional expectations, we show in the appendix that 
\begin{equation}\label{eq:dim_unbiased}
E[\hat\Delta_n] = E[Y_i(1) - Y_i(0)] = \Delta~.
\end{equation}
 We thus obtain that the estimator $\hat{\Delta}_n$ is also an unbiased estimator of average treatment effect in the super-population paradigm. Following a similar line of reasoning using the properties of conditional variances, we show in the appendix that
 \begin{equation}\label{eq:super_var}
\text{Var}[\hat\Delta_n] =  \frac{\text{Var}[Y_i(1)]}{n_1} + \frac{\text{Var}[Y_i(0)]}{n_0}~.
 \end{equation}
It is instructive to compare the limits of the variance expressions in \eqref{eq:fp_var} and \eqref{eq:super_var}.  To ensure nondegenerate limits, it is useful to scale the variances by $n$.  By doing so, we see that the limit of the super-population variance mirrors the limit of the finite-population variance in a regime where we sample a vanishing fraction of the total population, i.e., $n/N \rightarrow 0$.
Since the variance of the unit-level treatment effects does not appear in our expression for $\var[\hat{\Delta}_n]$ in \eqref{eq:super_var}, consistent variance estimation will be feasible in the super-population paradigm.


To discuss inference on $\Delta$, we document some necessary large-sample properties of $\hat{\Delta}_n$ in the super-population framework. Under the assumption that $E[Y_i^2(d)]<\infty$, it can be shown that 
\[\sqrt{n}(\hat{\Delta}_n - \Delta) \xrightarrow{d} N(0, V^{\rm cr})~,\]
as $n \rightarrow \infty$, where
\begin{equation} \label{eq:completerand-super}
   V^{\rm cr} = \frac{\text{Var}[Y_i(1)]}{\pi} + \frac{\text{Var}[Y_i(0)]}{1 - \pi}~,
\end{equation}
\citep[see, for instance,][]{bugni2018inference}. Using the above result, asymptotic inference on $\Delta$ is straightforward once we have an estimator of $V^{\rm cr}$. As in our discussion in Section \ref{sec:finpop_intro}, viewing $\hat{\Delta}_n$ as the estimator of the coefficient on $D_i$ obtained by the regression in \eqref{eq:OLS}, it can be shown that a consistent estimator of $V^{\rm cr}$ can be obtained from the resulting heteroskedasticity-robust variance estimator. An asymptotically valid $95\%$-confidence interval for $\Delta$ is therefore once again given by $C_n$ in \eqref{eq:Cn}.
Moreover, since this variance estimator is consistent, $C_n$ has \emph{exact} asymptotic coverage, that is
\[P\{\Delta \in C_n\} \rightarrow 0.95~,\]
as $n \rightarrow \infty$.

The above discussion illustrates an important feature of super-population and finite population analyses of randomized experiments: methods of inference developed in a super-population framework typically \emph{immediately} deliver conservative methods of inference from a finite population perspective.  While the preceding discussion was limited to completely randomized experiments, in Appendix \ref{sec:variance_heuristic}, we show that this feature holds much more generally under appropriate assumptions. With this in mind, the remainder of this article will focus on illustrating some of the major themes in the analysis of randomized experiments from the super-population perspective, but we will comment on any other important differences between these two perspectives whenever they arise.


\subsection{Further Reading}
Most of the material in this section is by-now standard, and several textbook treatments exist at various levels of formality: see in particular \cite{imbens2015causal}, \cite{athey2017econometrics}, and \cite{lehmann2022testing}. \cite{li2017general} provide formal statements and proofs of finite population central limit theorems. \cite{harshaw2021optimized} study a general method for constructing less conservative variance estimators in the design-based paradigm.


\section{Stratified Randomized Experiments} \label{sec:sbr}
In this section, we outline some benefits of \emph{stratification} and its consequences on subsequent experimental analyses. In a stratified randomized experiment, individuals are first divided into groups (i.e., strata) sharing similar values of their baseline covariates and then assigned to treatment so as to achieve ``balance" across the treatment and control groups: often, this amounts to simply performing complete randomization within each stratum. Stratification is extremely common in the design of randomized experiments in all parts of the sciences \citep[some examples in economics include][]{duflo2015education,berry2018impact,dizon-ross2019parents,callen2020data}. A primary motivation for stratification, going back to the work of \cite{fisher1935design}, is to ensure that the treatment and control groups are similar \emph{in the sample}, in contrast to complete randomization which can only ensure that this will be true \emph{in expectation}. As we will show, this property of stratification can lead to an increase in precision of the difference-in-means estimator $\hat{\Delta}_n$ relative to complete randomization, and as a result subsequent inferences will be unnecessarily conservative unless this is taken into consideration. In Section \ref{sec:strat_benefits} we illustrate the benefits of stratification via a simple example. In Section \ref{sec:strat_inference} we discuss inference in stratified randomized experiments, including inference with ``small" strata.

\subsection{Some Benefits of Stratification}\label{sec:strat_benefits}
\begin{figure}
\usetikzlibrary{positioning, shapes.geometric}

\begin{tikzpicture}
[
    node distance=1cm and 0.5cm,
    mynode/.style={draw,ellipse,align=center,minimum width=2.5cm, minimum height=1cm},
    myarrow/.style={-latex},
    mylabel/.style={font=\scriptsize},
    myrect/.style={draw,rectangle,minimum width=1cm,minimum height=0.5cm,align=center}
]
    \node[mylabel] at (0.5,4) (method) {\large Method of Randomization};
    \node[mylabel] at (-3,3) (complete) {\large Complete}; 
    \node[mylabel] at (4,3) (stratified) {\large Stratified}; 

    \node[mynode] (x0group) at (-3,2) {40};
    \node[myrect, below left=of x0group] (x0treat) {30\\treatment};
    \node[myrect, below right=of x0group] (x0control) {10\\control};
    \draw[myarrow] (x0group) -- (x0treat);
    \draw[myarrow] (x0group) -- (x0control);
    \node[mylabel, left=of x0group, xshift=-1.8cm, yshift=-0.5cm] (x0label) {\large $X = 0$};

    \node[mynode] (x1group) at (-3,-1) {60};
    \node[myrect, below left=of x1group] (x1treat) {20\\treatment};
    \node[myrect, below right=of x1group] (x1control) {40\\control};
    \draw[myarrow] (x1group) -- (x1treat);
    \draw[myarrow] (x1group) -- (x1control);
    \node[mylabel, left=of x1group, xshift = -1.8cm, yshift=-0.5cm] (x1label) {\large$X = 1$};

    \node[mynode] (sx0group) at (4,2) {40};
    \node[myrect, below left=of sx0group] (sx0treat) {20\\treatment};
    \node[myrect, below right=of sx0group] (sx0control) {20\\control};
    \draw[myarrow] (sx0group) -- (sx0treat);
    \draw[myarrow] (sx0group) -- (sx0control);

    \node[mynode] (sx1group) at (4,-1) {60};
    \node[myrect, below left=of sx1group] (sx1treat) {30\\treatment};
    \node[myrect, below right=of sx1group] (sx1control) {30\\control};
    \draw[myarrow] (sx1group) -- (sx1treat);
    \draw[myarrow] (sx1group) -- (sx1control);
    
    \node[mylabel] at (-7.2,3.1) (stratumgroup) {\large Stratum};
\end{tikzpicture}
\caption{Assignment across treatment and control for one realization of (CR) vs (SBR)}
    \label{fig:strat_pic}
\end{figure}
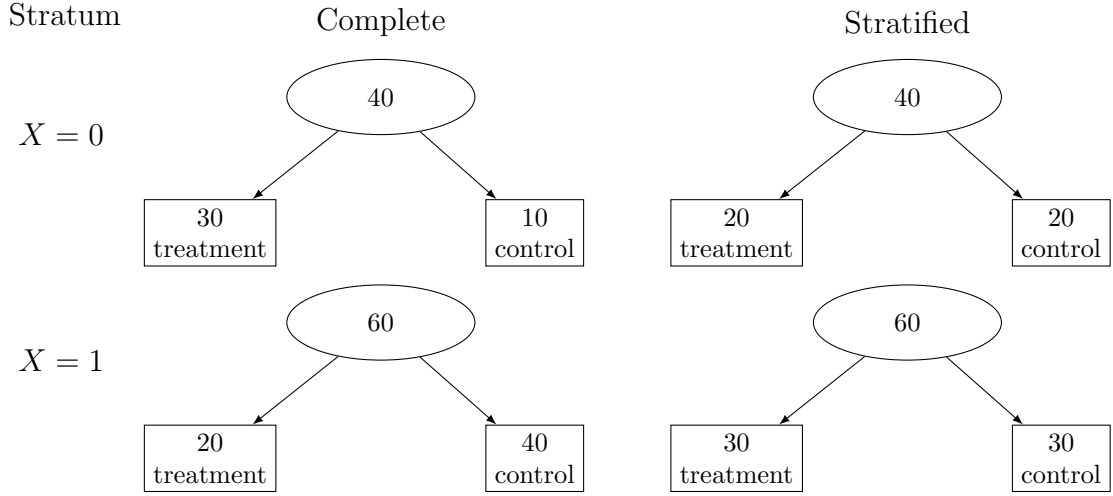

We begin by illustrating the benefits of stratification via a simple example. Suppose we have an experimental sample $\{(Y_i(1), Y_i(0), X_i): 1 \le i \le n\}$, where for now we assume $X_i \in \{0, 1\}$ are binary variables, and for convenience we assume $n$ is even. Consider the following two treatment assignment mechanisms for $D^{(n)}$:
\begin{enumerate}
    \item Complete Randomization (CR): Treatment is completely randomized with $\pi = \frac{1}{2}$ (see Section \ref{sec:setup} for a formal definition). 
    \item Stratified Block Randomization (SBR): Independently for each stratum (the sub-samples with $X_i = 0$ and $X_i = 1$), treatment is completely randomized with $\pi = \frac{1}{2}$. 
\end{enumerate}
Note that (SBR) as defined above uses the assignment proportion $\pi = \frac{1}{2}$ in both strata. Although maintaining a constant assignment proportion across all strata is the most common approach in practice, in principle we could also consider using different assignment proportions in each stratum, and we comment on the implications of doing so on subsequent analyses at the end of Section \ref{sec:strat_inference}. Although both assignment mechanisms (CR) and (SBR) assign exactly $n/2$ units to treatment, (CR) does not enforce that the \emph{composition} of the treated group across $X_i \in \{0, 1\}$ matches the composition in the total sample. For instance, suppose $n = 100$, where $40$ units have  $X_i = 0$ and $60$ units have $X_i = 1$. Figure \ref{fig:strat_pic} depicts one possible assignment that could result from employing (CR) versus (SBR). Although (CR) guarantees that exactly $50$ units are assigned to treatment, in this realization of the assignment, units with $X = 0$ are over-represented in the treatment group. In contrast, (SBR) reproduces the composition in the overall sample in both the treatment and control groups. 

To formalize this intuition, we analyze two different notions of bias for the difference-in-means estimator $\hat{\Delta}_n$ under both designs. To simplify the exposition, in this section, we perform our analyses conditional on the observable characteristics $X^{(n)}$ (although we emphasize that this does not materially change the conclusions). To that end, we will temporarily switch our parameter of interest to 
\[ \Delta_n(X^{(n)}) = \frac{1}{n} \sum_{1 \leq i \leq n} E[Y_i(1) - Y_i(0) | X_i]~, \]
which is the average effect of the treatment conditional on the covariates in the sample.

For the estimator $\hat \Delta_n$, we define the ex-ante bias as
\[ \mathrm{Bias}^{\rm ante}(X^{(n)}) = E[\hat \Delta_n | X^{(n)}] - \Delta_n(X^{(n)})~, \]
which averages over all possible realizations of the treatment assignments. This ex-ante bias measures the average bias obtained by repeatedly running the experiment on the same realization of the covariates. In contrast, we define the ex-post bias as the bias \emph{conditional on the realized treatment assignments}:
\[ \mathrm{Bias}^{\rm post}(X^{(n)}, D^{(n)}) = E[\hat \Delta_n | X^{(n)}, D^{(n)}] - \Delta_n(X^{(n)})~. \]

First we compare the ex-ante biases generated by (CR) and (SBR). Note that the marginal treatment probability of each unit satisfies $E[D_i|X^{(n)}] = \frac{1}{2}$ under both designs, i.e., the conditional probability that any given unit is assigned to treatment is one half. Combining this fact with the (conditional) exogeneity of treatment assignment, we obtain that
\[E[\hat{\Delta}_n|X^{(n)}] = \Delta_n(X^{(n)})~,\]
so that $\mathrm{Bias}^{\rm ante}(X^{(n)}) = 0$ for both designs. To compare the ex-post biases generated by (CR) and (SBR), consider the following decomposition:
\begin{align*}
   \mathrm{Bias}^{\rm post}(X^{(n)}, D^{(n)}) & = \frac{1}{n} \sum_{1 \leq i \leq n} (2D_i - 1) E[Y_i(1) + Y_i(0) | X_i] \\
   & = \frac{1}{n} E[Y_i(1) + Y_i(0) | X_i = 1] \cdot \mathrm{Imb}(1) + \frac{1}{n} E[Y_i(1) + Y_i(0) | X_i = 0] \cdot \mathrm{Imb}(0)~,
\end{align*}
where
\[ \mathrm{Imb}(x) = \# \{\text{treated units with } X_i = x\} - \# \{\text{untreated units with } X_i = x\} \]
is a measure of the \emph{imbalance} of treatment status for each possible value of the covariate $X_i$. By construction, (SBR) enforces that for \emph{any realization} of $D^{(n)}$ it is the case that $\mathrm{Imb}(x) = 0$. In contrast, as depicted in Figure \ref{fig:strat_pic}, this is not the case for (CR). As a consequence, it follows immediately that under (SBR), $\mathrm{Bias}^{\rm post}(X^{(n)}, D^{(n)}) \equiv 0$, whereas the ex-post bias under (CR) is not guaranteed to be identically zero unless $X_i$ is an irrelevant stratification variable in the sense that $E[Y_i(1) + Y_i(0)|X_i = 1] = E[Y_i(1) + Y_i(0)|X_i = 0]$. 


Next, we show that these properties of the ex-post bias have direct implications for the (ex-ante) \emph{variance} of $\hat{\Delta}_n$ under (CR) and (SBR). By the law of total variance:
\[ \var[\hat \Delta_n | X^{(n)}] = E[\var[\hat \Delta_n | X^{(n)}, D^{(n)}] | X^{(n)}] + \var[E[\hat \Delta_n | X^{(n)}, D^{(n)}] | X^{(n)}]~. \]
We show in the appendix that 
\begin{equation}\label{eq:EVarDelta}
E[\var[\hat \Delta_n | X^{(n)}, D^{(n)}] | X^{(n)}] = \frac{2}{n^2} \sum_{1 \leq i \leq n} (\var[Y_i(1) | X_i] + \var[Y_i(0) | X_i])~.
\end{equation}
As a result, $E[\var[\hat \Delta_n | X^{(n)}, D^{(n)}] | X^{(n)}]$ doesn't depend on the experimental design, and thus to compare $\var[\hat{\Delta}_n|X^{(n)}]$ under (CR) and (SBR) it suffices to study
\[ \var[E[\hat \Delta_n | X^{(n)}, D^{(n)}] | X^{(n)}] = E[\mathrm{Bias}^{\rm post}(X^{(n)}, D^{(n)})^2 | X^{(n)}]~. \]
In words, comparing $\var[\hat \Delta_n | X^{(n)}]$ under (CR) versus (SBR) amounts to a comparison of the second moment of the ex-post bias under both designs; because the ex-post bias is always zero under (SBR), the variance of $\hat \Delta_n$ is always smaller under (SBR) than (CR). In the next section, we discuss the implications of this increase in precision on subsequent inferences.

\subsection{Inference in Stratified Experiments}\label{sec:strat_inference}
In this section, we discuss the implications of stratification for inference on $\Delta$. For now, suppose that $X_i$ takes a finite number of values in $\mathcal X = \{1, 2, \ldots, |\mathcal{X}|\}$; this could be either because $X_i$ is naturally discrete (for instance, if $X_i$ denotes whether or not an individual graduated college), or because the researcher has discretized some continuous variables (for example, binning students by high or low test scores). In Section \ref{sec:small_strata} we consider settings where units are stratified as finely as possible based on potentially continuous covariates. To fix ideas, suppose the experiment was performed using stratified block randomization as defined in Section \ref{sec:strat_benefits}: independently in each sub-sample $X_i = x$, treatment is completely randomized with $\pi \in (0, 1)$.
Under appropriate assumptions, it can be shown that
\[\sqrt{n}(\hat{\Delta}_n - \Delta) \xrightarrow{d} N(0, V^{\rm sbr})~,\]
where 
\begin{equation}\label{eq:stratrand-super}
V^{\rm sbr} = \frac{\var[Y_i(1)]}{\pi} + \frac{\var[Y_i(0)]}{1 - \pi} - \pi(1-\pi)\var\left[E\left[\frac{Y_i(1)}{\pi}+\frac{Y_i(0)}{1-\pi}\Big|X_i\right]\right]~;
\end{equation}
see, for instance, \cite{bugni2018inference}. Comparing the variance $V^{\rm cr}$ obtained from complete randomization in \eqref{eq:completerand-super} to $V^{\rm sbr}$, we see that $V^{\rm sbr} \le V^{\rm cr}$ with equality only if $X$ is an irrelevant stratification variable in the sense that $E\left[\frac{Y_i(1)}{\pi} + \frac{Y_i(0)}{1-\pi}\Big|X_i\right]$ is constant; note that this is exactly the condition that guaranteed that the ex-post bias of $\hat{\Delta}_n$ under complete randomization with $\pi = 1/2$ was zero in Section \ref{sec:strat_benefits}.

Recall that we argued in Section \ref{sec:superpop_intro} that the heteroskedasticity-robust variance estimator from the regression described in \eqref{eq:OLS} is consistent for $V^{\rm cr}$ under complete randomization. It can be shown that the same holds under stratified block randomization, from which it follows that the robust variance estimator is generally \emph{conservative} for $V^{\rm sbr}$. In words, the robust variance estimator does not properly account for the gain in precision obtained by having performed stratification. How then could we conduct non-conservative inference on $\Delta$ in the presence of stratification? A straightforward option is to modify the variance estimator of $V^{\rm sbr}$ to account for the gain in precision. When the stratification variable $X$ takes only a finite number of values and there are at least two units in each of treatment and control, a simple solution is as follows: first, using the law of total variance, we re-write $V^{\rm sbr}$ as
\[V^{\rm sbr} = E\left[\frac{\var[Y_i(1)|X_i]}{\pi}\right] + E\left[\frac{\var[Y_i(0)|X_i]}{1 - \pi}\right] + \var[E[Y_i(1) - Y_i(0)|X_i]]~.\]
Exploiting the discreteness in $X$, we can expand this as
\[V^{\rm sbr} = \sum_{x \in \mathcal X}p(x)\left(\frac{\var[Y_i(1)|X_i = x]}{\pi} + \frac{\var[Y_i(0)|X_i = x]}{1 - \pi}\right) + \sum_{x \in \mathcal X} p(x)\left(E[Y_i(1) - Y_i(0)|X_i=x] - \Delta\right)^2~,\]
where $p(x) = P\{X_i = x\}$. From this expression it is clear how to construct a consistent estimator of $V^{\rm sbr}$ by simply replacing all of the unknown means and variances by their sample counterparts. 

Let us briefly comment on the implications of the above discussion for design-based inference. Although we argued at the end of Section \ref{sec:superpop_intro} that, in general, consistent estimators for the super-population variance immediately provide conservative estimators in the finite-population framework, it is worth mentioning that in this case the estimator is potentially ``excessively" conservative: indeed, from the design-based perspective, it can be shown that a less conservative estimator is given by any consistent estimator of
\begin{equation}\label{eq:strat_fp}
\sum_{x \in \mathcal X}\frac{n(x)}{n}\left(\frac{S^2_{1}(x)}{\pi} + \frac{S^2_{0}(x)}{1 - \pi}\right)~,
\end{equation}
where $n(x)$ denotes the number of observations in stratum $x$ and $S^2_{d}(x)$ for $d \in \{0, 1\}$ are the population variances of the potential outcomes in stratum $x$ \citep[see, for instance,][for details]{imbens2015causal}. Note that this expression mimics only the first component of $V^{\rm sbr}$; this is because the second component arises from the random fluctuations in $n(x)/n$ versus $p(x)$, but these exactly coincide in a design-based framework. However, because of this, consistent estimators of \eqref{eq:strat_fp} are \emph{not} guaranteed to be valid if we view the sample as being drawn from a larger (finite or super-) population.

Finally, we conclude with a short discussion on settings in which the assignment proportions differ across strata \citep[see, for instance,][for an example]{karlan2010expanding}. We continue to assume that treatment is assigned using stratified block randomization but we let $\pi(x)$, the fraction of units that are treated in stratum $x \in \{1, 2, \ldots, \mathcal X\}$, differ across strata. Our first observation is that the difference-in-means estimator is generally no longer consistent for $\Delta$; intuitively, since the probability of assignment differs across strata, the experimental design induces selection bias with respect to the stratification variable since the composition of the covariates differs between treatment and control. Instead, we could consider the following estimator which computes a weighted average of the stratum-specific difference-in-means estimators, with weights determined by the strata sizes:
\begin{equation} \label{eq:sat}
   \hat \Delta_n^{\rm sat} = \sum_{x \in \mathcal X} \frac{n(x)}{n} \hat \Delta_n(x)~, 
\end{equation}
where $n(x)$ denotes the number of observations in stratum $x$ and $\hat \Delta_n(x)$ denotes the difference-in-means estimator computed in stratum $x$. This estimator is consistent, and moreover it can be shown that its asymptotic variance is given by 
\[\sum_{x \in \mathcal X}p(x)\left(\frac{\var[Y_i(1)|X_i = x]}{\pi(x)} + \frac{\var[Y_i(0)|X_i = x]}{1 - \pi(x)}\right) + \sum_{x \in \mathcal X} p(x)\left(E[Y_i(1) - Y_i(0)|X_i=x] - \Delta\right)^2~,\]
for which a consistent estimator can once again be constructed by taking sample analogs; see \cite{bugni2019inference} for details.

\subsubsection{Variance Estimation with Small Strata}\label{sec:small_strata}
When $X$ was discrete, a natural estimator of $V^{\rm sbr}$ could be constructed by computing sample analogs of the means and variances of the potential outcomes at the stratum level. If, however, $X_i$ contains continuous components and the experiment is ``finely stratified" in such a way that there is only one treated or control observation per stratum, then this logic breaks down. A leading example of when this occurs is with pair-wise matching: in a matched pairs experiment, units are paired together based on their observed covariate values and then treatment is assigned such that, in each pair, one unit is selected at random to receive treatment and the other control. With only one observation per stratum with a given treatment, we cannot estimate the stratum-level variances by simply taking sample analogs as proposed in the previous section. As discussed in \cite{klar1997merits}, this added challenge to variance estimation has often been perceived as a fundamental analytical limitation of matched pair designs. In this section, we briefly illustrate how the variance of $\hat{\Delta}_n$ can still be consistently estimated by using a ``collapsed-strata" estimator in the spirit of \cite{hansen1953sample}. 

To streamline the exposition, in what follows we focus only on the setting where $\pi = \frac{1}{2}$ and units are matched into pairs. First, note that it can be shown under appropriate assumptions that, even with small strata, the limiting variance of $\hat{\Delta}_n$ is still given by \eqref{eq:stratrand-super} \citep[see, for instance,][]{bai2022inference}. The unconditional variances $\var[Y_i(1)]$ and $\var[Y_i(0)]$ in \eqref{eq:stratrand-super} are consistently estimable using their sample counterparts. We thus focus on estimating the last term on the right-hand side of \eqref{eq:stratrand-super}, which in this case is $(1/2)\var[E[Y_i(1) + Y_i(0) | X_i]]$. To that end, note that by elementary properties of the variance and the law of iterated expectations,
\[\var[E[Y_i(1) + Y_i(0) | X_i]] = E[E[Y_i(1) + Y_i(0) | X_i]^2] - E[Y_i(1) + Y_i(0)]^2~. \]
The second term on the right-hand side is again unconditional and thus easy to estimate using a sample analog. Therefore, it suffices to consistently estimate
\begin{equation} \label{eq:E^2}
   E[E[Y_i(1) + Y_i(0) | X_i]^2]~. 
\end{equation}
Intuitively, to estimate this last quantity we would want independent variation in $Y_i(1)$ and $Y_i(0)$ for each given value of $X_i$. Let us suppose for a moment that we did in fact have two pairs of units $\{Y_1, Y_2\}$ and $\{Y_3, Y_4\}$ sharing the same value of $X$. In each pair, there is exactly one treated and control observation per stratum. It follows that that $Y_1 + Y_2$ and $Y_3 + Y_4$ both share the same conditional mean $E[Y(1) + Y(0)|X]$. Moreover, across pairs, the outcomes are (conditionally) independent. As a consequence, we might conjecture that $(Y_1 + Y_2)(Y_3 + Y_4)$ is approximately equal to $E[Y(1) + Y(0)|X]^2 + (\epsilon_1 + \epsilon_2)(\epsilon_3+ \epsilon_4)$, where $E[\epsilon_i \epsilon_j | X^{(n)}] = 0$ for $i$ and $j$ in different pairs. Therefore, we can construct a consistent estimator for \eqref{eq:E^2} by averaging over these products across pairs of pairs. Of course in practice, this scenario is not realistic, as units which are paired together will typically not share the same value of $X$. If, however, matching is performed so that we expect pairs and ``adjacent" pairs to have similar characteristics, then this intuition can be formalized to construct a consistent variance estimator for \eqref{eq:E^2}, and thus it is possible to construct a consistent estimator of $V^{\rm sbr}$ even with ``small" strata; for a formal exposition of this idea, see \cite{bai2022inference,bai2024inference}.

\subsection{Further Reading}
The exposition in Section \ref{sec:strat_benefits} is most closely inspired by \cite{bai2022optimality}, who establishes the finite-sample optimality of certain matched-pair designs. Much of the discussion in Section \ref{sec:strat_inference} comes from \cite{bugni2018inference}, who study inference for covariate-adaptive randomization when the treated fraction is constant across strata; \cite{bugni2019inference} extends the analysis to settings with multiple treatments and where the treatment proportions are allowed to differ across strata. \cite{bai2022inference} studies inference for matched pair designs and develops the estimation strategy discussed in Section \ref{sec:small_strata}. \cite{cytrynbaum2023optimal} generalizes these estimation and inference procedures beyond pair-matching and jointly analyzes the stratification problem combined with the problem of selecting a representative sample based on covariates. \cite{bai2024inference} generalizes \cite{bai2022inference} to settings with multiple treatments. \cite{pashley2021insights} and \cite{bai2025new} provide overviews of design-based analyses of stratified experiments. 


\section{Regression Adjustment in Randomized Experiments}\label{sec:cov-adjust}
In this section, we consider the role of regression adjustment using baseline covariates in the analysis of randomized experiments. A primary motivation for regression adjustment is that it hopefully improves estimation precision in settings where the covariates are correlated with the experimental outcome. This practice has, however, often come under scrutiny.  An influential paper of \cite{freedman2008regression}, for instance, points out that standard linear regression adjustment only guarantees a gain in precision under strong assumptions, and concludes that in general ``[...] randomization does not justify the assumptions behind the OLS model." In Section \ref{sec:adjust-example} we review part of Freedman's critique, present a resolution popularized by \cite{lin2013agnostic}, and briefly discuss some implications for inference. In Section \ref{sec:adjust-general} we explain how Lin's resolution is a special case of an estimation procedure based on the doubly-robust moment condition of the average treatment effect, and show how this perspective leads to more general regression adjustment strategies beyond linear adjustment.

\subsection{Linear Regression Adjustment in a Completely Randomized Experiment}\label{sec:adjust-example}
We revisit the setting of a completely randomized experiment with $\pi \in (0, 1)$. In particular, we assume that treatment assignment is independent of the outcomes \emph{and} the baseline covariates; we briefly comment on settings where the treatment assignment may itself depend on the baseline covariates (for instance, in a stratified randomized experiment) in Section \ref{sec:adjust_inference} and provide further references in Section \ref{sec:further-adjust}. To begin, recall from \eqref{eq:OLS} that the unadjusted difference-in-means estimator can be described as the OLS estimator of the coefficient on $D_i$ in a linear regression of $Y_i$ on a constant and $D_i$. A natural starting point for regression adjustment is then to instead consider the OLS estimator of the coefficient on $D_i$ in a linear regression of $Y_i$ on a constant, $D_i$, and the covariates $X_i$:
\begin{equation}\label{eq:OLS_adjust}
\texttt{regress } Y_{i} \texttt{ on } \text{constant} +  D_i + X_i~.
\end{equation}
Let $\tilde \beta_n$ be the resulting estimator of the coefficient on $D_i$ from the above regression, and let $\tilde{\gamma}_n$ be the resulting estimator of the coefficient on $X_i$. We emphasize here that we do not view the regression in \eqref{eq:OLS_adjust} as describing the true data generating process for the outcomes $Y_i$, but treat this simply as the description of an estimation procedure. Since the experimental assignment guarantees that $D$ and $X$ are independent, it is not surprising given standard regression logic that $\tilde{\beta}_n$ remains consistent for the average treatment effect $\Delta$. One might expect further that $\tilde{\beta}_n$ is more efficient than the simple difference-in-means if $X_i$ is correlated with the experimental outcomes. It can be shown, however, that $\tilde{\beta}_n$ is asymptotically normal with variance given by \citep[see, for instance,][]{negi2021revisiting, ma2022regression}
\[V^{\rm pool} = V^{\rm cr} - \frac{1}{\pi(1 - \pi)}\gamma'\Sigma_{X}\gamma + \frac{2(2\pi - 1)}{\pi(1 - \pi)}\gamma'\Sigma_{X}(\gamma(1) - \gamma(0))~,\]
where $\gamma$ denotes the probability limit of $\tilde{\gamma}_n$, $\Sigma_X$ denotes the covariance matrix of $X_i$, and $\gamma(d)$ is the probability limit of the coefficient on $X_i$ in a linear regression of the \emph{potential outcome} $Y_i(d)$ on a constant and $X_i$. Note that the second term in this expression is always weakly negative, and is strictly negative whenever the covariates are correlated with the outcomes in the sense that $\gamma'\Sigma_X\gamma > 0$. However, the  sign of the third term in the variance is ambiguous, and as a result, there is no guarantee that $V^{\rm pool}$ is smaller than $V^{\rm cr}$ in general. Note that the third term of $V^{\rm pool}$ is zero in the special cases where either $\pi = 1/2$ or $\gamma(1) = \gamma(0)$, so that in these cases we can in fact conclude that $V^{\rm pool} \le V^{\rm cr}$. In words, regression adjustment based on the regression in \eqref{eq:OLS_adjust} is not guaranteed to improve precision in general, but (weakly) improves precision in the special cases where assignment is equal between treatment and control, or treatment effects are sufficiently homogeneous, e.g., if treatment effects are constant such that $Y_i(1) - Y_i(0) = \Delta$. 


The difficulty with estimation based on \eqref{eq:OLS_adjust} is that the regression pools together the observations under both treatment and control when estimating the relationship between the outcome and the covariates. To explain how this could be resolved, we first review the classical problem of estimation of a population mean using regression \cite[see, for example,][]{cochran1977sampling}. Let us suppose for a moment that $Y_i(d)$ and $E[X_i]$ were observed, and suppose we wished to estimate $E[Y_i(d)]$ using one of the following two regressions:
\begin{align}
    & \texttt{regress } Y_{i}(d) \text{ on } \text{constant} \label{eq:regd1} \\
    & \texttt{regress } Y_{i}(d) \text{ on } \text{constant} +  (X_i - E[X_i]) \label{eq:regd2}~.
\end{align}
Let $\hat{\alpha}^{(1)}_n$ be the resulting estimator of the coefficient on the constant from the regression in \eqref{eq:regd1}, and let $\hat{\alpha}^{(2)}_n$ and $\hat{\gamma}^{(2)}_n$ be the resulting estimators of the coefficient on the constant and the coefficient on $(X_i - E[X_i])$ from the regression in $\eqref{eq:regd2}$. We then obtain from elementary properties of regression that 
\[\hat{\alpha}^{(1)}_n = \frac{1}{n}\sum_{1 \le i \le n}Y_i(d)~,\]
and
\[\hat{\alpha}^{(2)}_n = \frac{1}{n}\sum_{1 \le i \le n}Y_i(d) - \hat{\gamma}^{(2)}_n\left(\frac{1}{n}\sum_{1 \le i \le n}(X_i - E[X_i])\right)~.\]
It follows that both $\hat{\alpha}_1$ and $\hat{\alpha}_2$ are consistent estimators of $E[Y_i(d)]$ (note here that it was important that we de-meaned the observable characteristics $X_i$), and that both are asymptotically normal with variances given by $\var[Y_i(d)]$ and 
\[\var[Y_i(d)] - \frac{\text{Cov}[Y_i(d),X]^2}{\var[X]}~,\]
respectively. We thus see immediately that $\hat{\alpha}_n^{(2)}$ is a more precise estimator of $E[Y_i(d)]$ whenever the covariates $X_i$ are correlated with the outcome $Y_i(d)$.

 \cite{lin2013agnostic} leverages this insight for the purpose of estimating the average treatment effect in a completely randomized experiment by effectively implementing the regression \eqref{eq:regd2} in each of the subsamples $D_i = d$ for $d \in \{0, 1\}$, and then taking the difference of the results. This procedure can be operationalized by estimating a linear regression model with an additional interaction term between $D$ and $X$:
 \begin{equation}\label{eq:OLS_interact}
\texttt{regress } Y_{i} \texttt{ on } \text{constant} +  D_i + X_i + D_i(X_i - \bar{X}_n)~.
\end{equation}
Let $\hat{\beta}_n$ be the resulting estimator of the coefficient on $D_i$ from the above regression. It can then be shown that $\hat{\beta}_n$ is a consistent and asymptotically normal estimator of $\Delta$ and that its asymptotic variance $V^{\rm sat}$ satisfies $V^{\rm sat} - V^{cr} \le 0$, with equality if and only if
\begin{equation} \label{eq:linear-irrelevant}
    \cov \left [\frac{Y_i(1)}{\pi} + \frac{Y_i(0)}{1 - \pi}, X_i \right ] = 0
\end{equation}
\citep[see, for instance,][]{negi2021revisiting, ma2022regression}. From this we can see that regression based on \eqref{eq:OLS_interact} always weakly improves precision regardless of the value of $\pi$, and we should expect that it strictly improves precision whenever the potential outcomes are correlated with the baseline covariates (outside of pathological cases). Further note that \eqref{eq:linear-irrelevant} holds if $E\left[\frac{Y_i(1)}{\pi} + \frac{Y_i(0)}{1-\pi}\Big|X_i\right]$ is constant; this is exactly the condition that guaranteed that $X_i$ is an irrelevant stratification variable in Section \ref{sec:sbr}. From this we can conclude that if $X_i$ is an irrelevant variable for stratification, then we cannot increase estimation efficiency through adjusting for $X_i$ either. We further explore the relationship between stratification and regression adjustment at the end of Section \ref{sec:adjust-general}. 

We conclude this section by noting that the above analysis suggests that, when $\pi = 1/2$, regression adjustment based on \eqref{eq:OLS_adjust} may be preferred to adjustment based on \eqref{eq:OLS_interact}; in this case both estimators have the same asymptotic variance and \eqref{eq:OLS_adjust} involves the estimation of fewer parameters. We caution, however, that this conclusion is very specific to the case of a binary treatment, and that strategies based on interacted models tend to generalize more broadly beyond the special case we considered here.

\subsubsection{Inference on $\Delta$ when using Linear Regression Adjustment: Some Caveats}\label{sec:adjust_inference}

We briefly discuss some complications surrounding inference on $\Delta$ based on the regressions \eqref{eq:OLS_adjust} and \eqref{eq:OLS_interact} under complete randomization, and mention some related complications when generalizing beyond complete randomization. Recall that in Section \ref{sec:fin/super} we explained that the robust variance estimator obtained from the regression in \eqref{eq:OLS} is consistent for the asymptotic variance of the difference-in-means estimator $\hat{\Delta}_n$. A natural follow-up question is then whether or not the robust variance estimators obtained from the regressions \eqref{eq:OLS_adjust} and \eqref{eq:OLS_interact} are consistent for (or at least an upper bound on) the asymptotic variances of $\tilde{\beta}_n$ and $\hat{\beta}_n$, respectively. In the case of the regression in \eqref{eq:OLS_adjust}, the answer is yes: the robust variance estimator is consistent because $\tilde{\beta}_n$ is simply the OLS estimator of the best linear predictor of $Y$ given $1, D,$ and $X$. In the case of the regression in \eqref{eq:OLS_interact}, however, the answer is \emph{no} whenever we view the sample as being drawn from a larger (finite or super-) population.  The issue here is similar to our discussion surrounding consistent estimators of the variance when studying stratified randomized experiments in Section \ref{sec:strat_inference}. In particular, note that the regression in \eqref{eq:OLS_interact} involves a de-meaning of the baseline covariates, and when we view the sample as being drawn from a larger population, the random fluctuations in $\bar{X}_n$ versus $E[X_i]$ need to be taken into consideration. As a result, the robust variance estimator obtained from regression \eqref{eq:OLS_interact} is \emph{not} guaranteed to be valid outside of the design-based framework (in which case $\bar{X}_n$ is non-random). When moving beyond complete randomization to settings with stratified assignment, the complications become more subtle. When stratification is based on a finite number of discrete categories, an easy solution is to perform regression adjustment in each stratum separately and then aggregate the stratum-level estimates; we provide relevant references for this and more complicated settings involving stratification in Section \ref{sec:further-adjust}. With these considerations in mind, we suggest that practitioners use our {\tt R} command {\tt sreg} to perform regression adjustment in (stratified) randomized experiments: see \cite{trifonov2025sreg}.

\subsection{Double Robustness and General Regression Adjustments}\label{sec:adjust-general}
In this section, we explain how Lin's interacted regression \eqref{eq:OLS_interact} is a special case of a more general class of estimators based on a \emph{doubly-robust} moment function. To simplify the exposition we once again assume that treatment is completely randomized with treated fraction $\pi \in (0, 1)$. Note that because the treatments are independent of the potential outcomes and covariates, the probability of assignment of any individual $i$ satisfies $P \{D_i = 1 | X_i = x\} \equiv \pi$. Moreover, it follows that the treatments are independent of the potential outcomes \emph{conditional} on the covariates.


Consider the parameter $\Delta_0$ defined by the following moment equation:
\begin{equation} \label{eq:doublyrobust}
   E\left[ \frac{D_i (Y_i - \tilde{\mu}_1(X_i))}{\pi} - \frac{(1 - D_i) (Y_i - \tilde{\mu}_0(X_i))}{1 - \pi} + \tilde{\mu}_1(X_i) - \tilde{\mu}_0(X_i) - \Delta_0 \right] = 0~,
\end{equation}
where $\tilde{\mu}_1(x), \tilde{\mu}_0(x)$ are arbitrary researcher-defined functions of $x$ which we call the ``working models'' for the conditional expectations $E[Y_i(1) | X_i = x]$ and $E[Y_i(0) | X_i = x]$. Note that
\begin{align*}
    E \left [ \frac{D_i (Y_i - \tilde{\mu}_1(X_i))}{\pi} + \tilde{\mu}_1(X_i) \right ]  & = E \left [ \frac{D_i Y_i(1)}{\pi} - \frac{D_i \tilde{\mu}_1(X_i)}{\pi} + \tilde{\mu}_1(X_i) \right ] \\
    & = E \left [ E \left [ \frac{D_i Y_i(1)}{\pi} \bigg | X_i \right ] - \frac{E[D_i | X_i] \tilde{\mu}_1(X_i)}{\pi} + \tilde{\mu}_1(X_i) \right ] \\
    & = E \left [\frac{\pi E[Y_i(1) | X_i]}{\pi}  - \frac{\pi \tilde{\mu}_1(X_i)}{\pi} + \tilde{\mu}_1(X_i) \right ] \\
    & = E[Y_i(1)]~.
\end{align*}
 Combining this with a similar argument for the other symmetric term in \eqref{eq:doublyrobust}, it follows that the solution to  \eqref{eq:doublyrobust} is $\Delta_0 = \Delta$ for any choice of working models $\tilde{\mu}_1(\cdot)$ and $\tilde{\mu}_0(\cdot)$.  Equation \eqref{eq:doublyrobust} is the famous ``doubly-robust" moment equation for the ATE due to \cite{robins1995analysis}.\footnote{The term ``doubly-robust" is due to the fact that, if conversely $\tilde{\mu}_1(x) = E[Y_i(1)|X_i=x]$ and $\tilde{\mu}_0(x) = E[Y_i(0)|X_i=x]$, equation \eqref{eq:doublyrobust} still identifies $\Delta$ even if $\pi \ne P\{D_i = 1|X_i =x \}$.} Since $\pi$ is known in a randomized experiment by construction, estimators based on taking a sample analog of $\eqref{eq:doublyrobust}$ will be consistent for $\Delta$ regardless of the choice of $\tilde{\mu}_1(\cdot)$ and $\tilde{\mu}_0(\cdot)$. 
In this way, we obtain the well-known augmented inverse-propensity weighted (AIPW) estimator of $\Delta$:
\begin{equation} \label{eq:covadj}
    \hat{\Delta}^{\rm AIPW}_n = \frac{1}{n} \sum_{1 \leq i \leq n} \Big ( \frac{D_i (Y_i - \hat \mu_{1, n}(X_i))}{\pi} - \frac{(1 - D_i) (Y_i - \hat \mu_{0, n}(X_i))}{1 - \pi} + \hat \mu_{1, n}(X_i) - \hat \mu_{0, n}(X_i) \Big )~,
\end{equation}
where $\hat{\mu}_{1,n}(\cdot)$ and $\hat{\mu}_{0,n}(\cdot)$ are consistent estimators of the working models. $\hat{\Delta}^{\rm AIPW}_n$ can recover a wide range of covariate-adjusted estimators of $\Delta$ by specifying different choices of $\tilde{\mu}_1(\cdot)$ and $\tilde{\mu}_0(\cdot)$ and their corresponding estimators. For instance, if we set $\hat{\mu}_{d,n} = \frac{1}{n_d}\sum_{1 \le i \le n}Y_iI\{D_i = d\}$, then we recover the difference-in-means estimator $\hat{\Delta}_n$ (if we instead set $\hat{\mu}_{1,n}(\cdot) = \hat{\mu}_{0,n}(\cdot) = 0$, then we recover the Horvitz-Thompson estimator of the average treatment effect). Alternatively, if we let $\hat{\gamma}_n(d)$ be the OLS estimator of the coefficient on $X_i$ in a regression of $Y_i$ on a constant and $X_i$ for observations with $D_i = d$, and set
\begin{align*}
    \hat \mu_{d, n}(X_i) & = (X_i - \bar X_n)' \hat \gamma_n(d)~,
\end{align*}
for $d \in \{0, 1\}$, then $\hat{\Delta}^{\rm AIPW}_n$ recovers Lin's interacted linear regression estimator \eqref{eq:OLS_interact}. 

If $\hat{\mu}_{1,n}(\cdot)$ and $\hat{\mu}_{0,n}(\cdot)$ are appropriately chosen non-parametric estimators of the true conditional mean functions $E[Y_i(1) | X_i = x]$ and $E[Y_i(0) | X_i = x]$, then $\hat{\Delta}^{\rm AIPW}_n$ becomes a \emph{non-parametric} regression-adjusted estimator of the average treatment effect. Under appropriate assumptions, it can be shown that $\hat{\Delta}^{\rm AIPW}_n$ is asymptotically normal with variance given by 
\[V^\ast = E\left[\frac{\var[Y_i(1)|X_i]}{\pi}\right] + E\left[\frac{\var[Y_i(0)|X_i]}{1 - \pi}\right] + \var[E[Y_i(1) - Y_i(0)|X_i]]~,\]
see \cite{tu2023unified}, \cite{rafi2023efficient} for details. Importantly, this variance is the efficient variance for estimating $\Delta$ using a randomized experiment when the probability of assignment is exogenously constrained to be $\pi$ \citep[see][]{armstrong2022asymptotic,rafi2023efficient,bai2025efficiency}. 

Note that $V^\ast$ coincides exactly with the asymptotic variance $V^{\rm sbr}$ obtained when using the unadjusted estimator $\hat{\Delta}_n$ in a finely stratified randomized experiment with ``small" strata (see Section \ref{sec:small_strata}). We have thus demonstrated two alternative methods for achieving the efficient variance $V^*$ when estimating $\Delta$ via a randomized experiment:\footnote{Of course, this list of methods is not exhaustive. For instance, efficient estimation could also be achieved under i.i.d.\ randomization using a standard inverse-propensity weighted (IPW) estimator, where the propensity score is \emph{non-parametrically} estimated as a function of $X$: see in particular \cite{hirano2003efficient} for details.} 
\begin{enumerate}
    \item Assign treatment using complete randomization and estimate $\Delta$ using $\hat{\Delta}_n^{\rm AIPW}$ with suitable non-parametric estimators $\hat{\mu}_{d,n}(\cdot)$ of the conditional means $E[Y_i(d)|X_i]$.
    \item Assign treatment using ``fine stratification" (for instance, if $\pi = 1/2$ this could be a matched pairs design as described in Section \ref{sec:small_strata}) and estimate $\Delta$ using $\hat{\Delta}_n$.
\end{enumerate} 
This demonstrates that experiments which assign treatment using fine stratification effectively perform non-parametric regression adjustment ``\emph{by design}"; this feature of finely stratified experiments is further explored in \cite{cytrynbaum2023optimal} and \cite{bai2025efficiency}. In principle, both strategies could be combined, by first stratifying on the covariates which might be expected to be most predictive for the outcome of interest, and then adjusting for a (potentially high-dimensional) set of additional covariates via regression ex-post: see \cite{bai2023covariate} for details.

\subsection{Further Reading}\label{sec:further-adjust}
The exposition in Section \ref{sec:adjust-example} closely follows \cite{negi2021revisiting} and \cite{ma2022regression}, although some of their expressions have been modified so that they could be more easily related to Sections \ref{sec:fin/super} and \ref{sec:sbr} of this paper.  The use of doubly-robust estimators for regression adjustment in randomized experiments goes back at least to the work of \cite{tsiatis2008covariate}; \cite{tu2023unified} and \cite{rafi2023efficient} extend these results to general stratified experiments with finitely many strata. \cite{bai2023covariate} and \cite{cytrynbaum2023covariate} consider regression adjustment in settings with ``small" strata, in the sense of Section \ref{sec:small_strata}.  \cite{wang2023model} study regression adjustment for general parameters defined by estimating equations. Recent work on design-based analyses of regression adjusted estimators can be found in \cite{aronow2013class}, \cite{wu2018loop}, \cite{liu2021randomization}, \cite{chang2023design}, \cite{chiang2023regression}. 

\section{The Analysis of Cluster Randomized Experiments}\label{sec:clust}
Until now, our focus has been on experiments where treatment is assigned at the individual level and each individual's outcome depends only on their own treatment. In this section, we consider randomized experiments where treatment is assigned at an aggregated level which we call a \emph{cluster}: for example, when evaluating an educational intervention we may observe outcomes at the student-level, but assign treatment at the school-level, so that every student in the school receives the same treatment. This type of cluster-level assignment is extremely common in practice: see, for instance, \cite{angrist2009effects, banerjee2015miracle, crepon2015estimating, bruhn2016impact, romero2020outsourcing}. There are two common explanations for why a researcher would consider cluster-level assignment as opposed to individual-level assignment: first, there could be logistical constraints on the experiment that require that every individual in a cluster receives the same treatment. Second, we may be concerned that there is treatment \emph{interference}, i.e., the treatment statuses of individuals in a cluster may affect the outcomes of others. In Section \ref{sec:clust_params} we introduce the framework and define some relevant analogs to the average treatment effect in this setting. In Section \ref{sec:clust_analysis}, we discuss inference in cluster randomized experiments under complete randomization and stratified block randomization. 

\subsection{Defining Average Treatment Effects in Cluster Randomized Experiments}\label{sec:clust_params}
To accommodate cluster-level assignment, we first modify our notation relative to what we have considered thus far. Let $Y_{i,g}$ denote the observed outcome of the $i$th unit in the $g$th cluster, $D_g$ denote an indicator for whether or not the $g$th cluster is treated, and $N_g$ the size of the $g$th cluster.  Further denote by $Y_{i,g}(1)$ the potential outcome of the $i$th unit in the $g$th cluster if the cluster is treated and by $Y_{i,g}(0)$ the potential outcome of the $i$th unit in the $g$th cluster if not treated. As usual, the observed outcome and potential outcomes are related to treatment assignment by the relationship 
\begin{equation*} 
Y_{i,g} = Y_{i,g}(1)D_g + Y_{i,g}(0)(1 - D_g)~.
\end{equation*}
In practice it is sometimes the case that the researcher does not sample all of the units in a given cluster. To allow for this possibility, define $\mathcal{M}_g$ to be the subset of $\{1, \ldots, N_g\}$ corresponding to the observations within the $g$th cluster that are sampled by the researcher. For example, in the event that all observations in a cluster are sampled, $\mathcal{M}_g = \{1, \ldots, N_g\}$ and $|\mathcal{M}_g| = N_g$.

With this updated notation, our sampling framework models 
\[\{((Y_{i,g}(1),Y_{i,g}(0) : 1 \le i \le N_g),\mathcal{M}_g, N_g) : 1 \leq g \leq G \}~,\] 
as a collection of $G$ independent and identically distributed draws from a distribution of clusters. Importantly, we note here that the cluster sizes $N_g$  are modeled as random variables which are potentially related to the potential outcomes. In this framework, we introduce two natural parameters that arise as generalizations of the average treatment effect $\Delta$ which we focused on in earlier sections. These parameters differ in the way they aggregate, or average, the individual level treatment effects.

Both parameters of interest we consider can be written as weighted averages of the cluster-level average treatment effects: 
\begin{equation*} 
E\left [\omega_g \left ( \frac{1}{N_g} \sum_{1 \leq i \leq N_g} \left(Y_{i,g}(1) - Y_{i,g}(0)\right) \right ) \right ] ~,
\end{equation*}
for different choices of (possibly random) weights $\omega_g$ satisfying $E[\omega_g] = 1$.  The first parameter of interest corresponds to the choice of $\omega_g = 1$, thus weighting the average effect of the treatment across clusters equally:
\begin{equation*} 
\Delta^{\rm eq} := E\left [ \frac{1}{N_g} \sum_{1 \leq i \leq N_g} \left(Y_{i,g}(1) - Y_{i,g}(0)\right) \right ]~.
\end{equation*}
We refer to this quantity as the equally-weighted cluster-level average treatment effect. Since $\Delta^{\rm eq}$ assigns an equal weight to each cluster regardless of size, it can be thought of as the average treatment effect where the clusters themselves are the units of interest. The second parameter of interest corresponds to the choice of $\omega_g = N_g/E[N_g]$, thus weighting the average effect of the treatment across clusters in proportion to their size:
\begin{equation*} 
\Delta^{\rm size} := E\left [ \frac{1}{E[N_g]} \sum_{1 \leq i \leq N_g} \left(Y_{i,g}(1) - Y_{i,g}(0)\right) \right ]~.
\end{equation*}
We refer to this quantity as the size-weighted cluster-level average treatment effect. Since $\Delta^{\rm size}$ assigns a weight proportional to each cluster's size, it can be thought of as the average treatment effect where individuals are the units of interest.

Note that in empirical settings with treatment effect heterogeneity (so that $Y_{i,g}(1) - Y_{i,g}(0)$ is random) and cluster-size heterogeneity, we should expect that $\Delta^{\rm eq}$ and $\Delta^{\rm size}$ are indeed distinct parameters with differing policy interpretations. For instance, suppose the experiment studies the effect of an educational intervention on students' reading level. If the policy-maker is interested in raising the average reading level across all students, then the magnitude of $\Delta^{\rm size}$ is the relevant parameter. If, on the other hand, the policy-maker has concerns about raising the average reading level across all schools, then the magnitude of $\Delta^{\rm eq}$ would also be important to consider.

\subsection{Inference in Cluster Randomized Experiments}\label{sec:clust_analysis}
In this section, we study estimation and inference on $\Delta^{\rm eq}$ and $\Delta^{\rm size}$. We begin with the setting of a completely randomized experiment; note that in the context of a cluster randomized experiment this means that a fraction $\pi \in (0, 1)$ of \emph{clusters} is assigned to treatment and the rest to control. 

We begin by first studying the probability limit of the difference-in-means estimator obtained from a regression of the individual-level outcomes $Y_{i,g}$ on a constant and the cluster-level treatment $D_g$. With our new notation, this is given by
\[\hat \Delta_{G} := \frac{\sum_{1 \leq g \leq G} \sum_{i \in \mathcal{M}_g} Y_{i,g} D_g}{\sum_{1 \leq g \leq G} |\mathcal{M}_g| D_g} - \frac{\sum_{1 \leq g \leq G} \sum_{i \in \mathcal{M}_g} Y_{i,g} (1- D_g)}{\sum_{1 \leq g \leq G} |\mathcal{M}_g| (1- D_g)}~.\]
It can be shown under appropriate assumptions that 
\[\hat \Delta_{G} \stackrel{P}{\rightarrow} E\left [ \frac{1}{E[|\mathcal{M}_g|]} \sum_{i \in \mathcal{M}_g} \left(Y_{i,g}(1) - Y_{i,g}(0)\right) \right ] =: \vartheta~,\] 
as $G \rightarrow \infty$; see \cite{bugni2022inference} for details. This parameter corresponds to a \emph{sample}-weighted cluster-level average treatment effect and in general does not coincide with either $\Delta^{\rm eq}$ or $\Delta^{\rm size}$. Of course, in some special cases, $\vartheta$ does coincide with $\Delta^{\rm eq}$ or $\Delta^{\rm size}$. For instance, if we sample the same number of observations in every cluster, then $\vartheta$ coincides with $\Delta^{\rm eq}$. If instead we sample observations proportionally to the size of each cluster, then $\vartheta$ coincides with $\Delta^{\rm size}$.

In order to do inference on $\Delta^{\rm eq}$ and $\Delta^{\rm size}$ regardless of the specific choice of sampling design, we now present estimators which are consistent for these parameters more generally. In the case of $\Delta^{\rm eq}$, we consider the following difference-in-means estimator computed on the cluster-average outcomes:
\[\hat \Delta^{\rm eq}_{G} := \frac{\sum_{1 \leq g \leq G} \bar Y_g D_g}{\sum_{1 \leq g \leq G} D_g} - \frac{\sum_{1 \leq g \leq G} \bar Y_g (1- D_g)}{\sum_{1 \leq g \leq G} (1- D_g)}~,\]
where $\bar{Y}_g = \frac{1}{|\mathcal{M}_g|}\sum_{i \in \mathcal{M}_g} Y_{i,g}$. Note that $\hat \Delta^{\rm eq}_G$ may be obtained as the estimator of the coefficient on $D_g$ from the following regression:
\begin{equation}\label{eq:OLS_cluster}
\texttt{regress } \bar{Y}_{g} \texttt{ on } \text{constant} +  D_g~;
\end{equation}
as such, it is exactly the difference-in-means estimator obtained from viewing the clusters as the experimental units of interest, with outcomes defined by their cluster averages. Under appropriate assumptions it can be shown that
\[\sqrt G (\hat \Delta^{\rm eq}_{G} - \Delta^{\rm eq}) \stackrel{d}{\rightarrow} N(0,V^{\rm eq})~,\]
as $G \rightarrow \infty$, where 
\[
V^{\rm eq} = \frac{{\var}[\bar Y_g(1)]}{\pi}  + \frac{{\var}[\bar Y_g(0)]}{1-\pi}~,
\]
with $\bar{Y}_g(d) = \frac{1}{|\mathcal{M}_g|}\sum_{i \in \mathcal{M}_g} Y_{i,g}(d)$ \citep[see][for details]{bugni2022inference}. We thus obtain that $\hat \Delta^{\rm eq}_G$ is a consistent and asymptotically normal estimator of $\Delta ^{\rm eq}$ with asymptotic variance which exactly mirrors the asymptotic variance $V^{\rm cr}$ obtained in Section \ref{sec:superpop_intro}, but with the individual level outcomes $Y_i(d)$ replaced with the cluster-level average outcomes $\bar{Y}_g(d)$. It is therefore not surprising that a consistent estimator of $V^{\rm eq}$ can be obtained from the resulting heteroskedasticity-robust variance estimator of the regression in \eqref{eq:OLS_cluster}. More generally, when studying $\Delta^{\rm eq}$ in a cluster randomized experiment, all of the tools introduced in Sections \ref{sec:fin/super}--\ref{sec:cov-adjust} can be applied by simply studying the experiment as if the clusters were individuals with outcomes given by the cluster averages $\bar{Y}_g$. 

In the case of $\Delta^{\rm size}$, we consider the following cluster-size weighted difference-in-means estimator:
\[
\hat \Delta^{\rm size}_G = \frac{\sum_{1 \leq g \leq G} \bar Y_g N_g D_g}{\sum_{1 \leq g \leq G} N_g D_g} - \frac{\sum_{1 \leq g \leq G} \bar Y_g N_g (1- D_g)}{\sum_{1 \leq g \leq G} N_g (1- D_g)}~,
\]
Note that $\hat \Delta^{\rm size}_G$ may be obtained as the estimator of the coefficient on $D_g$ in the following \emph{weighted} least squares regression:
\begin{equation}\label{eq:WLS_clust} \texttt{regress } Y_{i,g} \texttt{ on } \text{constant} +  D_g \texttt{ using weights } {N_g/|\mathcal{M}_g|}~.
\end{equation}
Under appropriate assumptions it can be shown that
\[\sqrt G (\hat \Delta^{\rm size}_G - \Delta^{\rm size}) \stackrel{d}{\rightarrow} N(0,V^{\rm size})~,\] as 
$G \rightarrow \infty$, where
\begin{equation*}
V^{\rm size} := \frac{{\var}[\widetilde Y_g(1)]}{\pi}  + \frac{ {\var}[\widetilde Y_g(0)]}{1-\pi}~,
\end{equation*}
with
\begin{eqnarray*}
\widetilde{Y}_g(d) &:=& \frac{N_g}{E[N_g]}\left(\bar{Y}_g(d) - \frac{E[\bar{Y}_g(d)N_g]}{E[N_g]}\right)~,
\end{eqnarray*}
see \cite{bugni2022inference} for details. We thus obtain that $\hat \Delta ^{\rm size}_G$ is a consistent and asymptotically normal estimator of $\Delta ^{\rm size}$ with asymptotic variance which again mirrors $V^{\rm cr}$, but now with the individual-level outcomes $Y_i(d)$ replaced with the \emph{re-scaled} cluster-level outcomes $\tilde{Y}_g(d)$. It turns out that in this case, a consistent estimator of $V^{\rm size}$ can be obtained from the regression \eqref{eq:WLS_clust} by computing the resulting \emph{cluster-robust} variance estimator. Equivalently, it is straightforward to construct a sample analog estimator of $V^{\rm size}$ where the infeasible re-scaled outcomes $\widetilde{Y}_g(d)$ are replaced by feasible analogs:
\[\hat{Y}_g :=  \frac{N_g}{\frac{1}{G}\sum_{1 \le j \le G}N_j}\left(\bar{Y}_g - \frac{\frac{1}{G}\sum_{1 \le j \le G}\bar{Y}_jI\{D_j = D_g\}N_j}{\frac{1}{G}\sum_{1 \le j \le G}I\{D_j = D_g\}N_j}\right)~.\] 
We find this latter approach particularly useful when moving beyond completely randomized experiments to settings with stratification: for example, under stratified block randomization, the limiting variance of $\hat \Delta_G^{\rm size}$ equals
\[ \frac{\var[\widetilde Y_g(1)]}{\pi} + \frac{\var[\widetilde Y_g(0)]}{1 - \pi} - \pi(1-\pi)\var\left[E\left[\frac{\widetilde Y_g(1)}{\pi}+\frac{\widetilde Y_g(0)}{1-\pi}\Big|X_g\right]\right]~, \]
where $X_g$ denotes the variables used in stratification \citep[see][for details]{bugni2022inference, bai2023inference-1}. This variance mirrors the variance $V^{\rm sbr}$ for individual-level stratified experiments defined in \eqref{eq:stratrand-super}, and therefore similar inference procedures apply to cluster randomized experiments by modifying the individual-level outcomes to suitable cluster-level counterparts such as $\hat Y_g$. 

\subsection{Further Reading}
The material in this section is most closely related to \cite{bugni2022inference} and \cite{bai2023inference-1}.
\cite{donner2000design} contains a textbook treatment of early work in cluster randomized experiments. \cite{su2021model} and \cite{wang2024model} study regression adjustment for cluster randomized experiments without stratification (\cite{bugni2022inference} and \cite{bai2023inference-1} discuss extensions of these results to settings with stratification).
Other recent work on cluster randomized experiments (mostly from a design-based perspective) includes \cite{imai2009essential},  \cite{schochet2013estimators}, \cite{middleton2015unbiased}, \cite{schochet2021design}, \cite{de_chaisemartin2024at}.

\section{Other Topics}\label{sec:other}
\subsection{Treatment Effect Heterogeneity and Quantile Treatment Effects} \label{sec:qtehte}
This article focused exclusively on estimation and inference of the unconditional average treatment effect. Recent work on the analysis of randomized experiments has studied inference for quantile treatment effects: see in particular \cite{zhang2020quantile}, \cite{jiang2021bootstrap}, and \cite{jiang2023regression}. We may also be interested in heterogeneity of the average treatment effect as a function of the observable characteristics (i.e., the conditional average treatment effect, or CATE). Although estimation and inference on the CATE has been an extremely active area of research in causal inference more broadly \citep[see, for instance,][for an in-depth discussion and further references]{kennedy2023towards}, to our knowledge almost all of this work maintains the assumption that treatment assignment is independent and identically distributed across individuals; this precludes, for instance, stratified block randomization or pair matching. Some exceptions are \cite{zhang2023interaction}, who present tests for treatment-covariate interactions and study their properties under general covariate-adaptive stratified randomization schemes, and \cite{ding2019decomposing, cytrynbaum2025finely}, who study parametric approximations to the CATE under complete randomization and finely stratified re-randomization designs, respectively.

\subsection {Re-randomization}
Re-randomization is a method of experimental assignment which, in a similar spirit to stratification, is intended to enforce ``balance" in the covariate distributions between treatment and control groups. In a standard re-randomization procedure, researchers specify a balance criterion for the covariate values and then repeatedly generate assignments using a completely randomized design until an assignment is found which achieves an acceptable covariate distribution according to the balance criterion. An excellent historical summary is given in \cite{morgan2012rerandomization}. Similarly to stratification, re-randomization leads to an increase in precision relative to complete randomization, and as a result inferences will be unnecessarily conservative unless this is taken into consideration. However, unlike with stratification, where correcting this issue usually amounts to simply modifying the standard errors, corrected inferences for re-randomization are sometimes further complicated by the fact that the limiting distribution of the difference-in-means estimator is not asymptotically normal \citep[although normality can be restored using ex-post linear adjustment: see][for details]{li2020rerandomization-2,cytrynbaum2025finely}. For an in-depth theoretical discussion of re-randomization, see  \cite{li2018asymptotic}, \cite{li2020rerandomization}, \cite{zhao2021no}, \cite{lu2023design}, \cite{branson2024power}. \cite{wang2023rerandomization} and \cite{cytrynbaum2025finely} study how to combine re-randomization with stratification.

\subsection{Multiple Testing}

Our discussion has so far mostly focused on inference about a single parameter of interest, namely the average effect of a binary treatment on an outcome of interest.  In many experiments, however, there may be many parameters of interest: there may be multiple treatments, and so it may be of interest to compare the average effect of each of these treatments with the control or with each other; there may be multiple outcomes of interest, and so it may be of interest to examine these effects for each of these outcomes of interest; finally, there may be multiple subgroups of interest (defined by observed, baseline characteristics, as in Section \ref{sec:qtehte}), and so it may be of interest to examine these effects separately for these different subgroups.  In many cases, the methods described previously can be modified in a straightforward fashion for inference about any one of these parameters, but it is often of interest to examine at least some subset of them simultaneously in order to determine, e.g., which of the parameters are equal to zero or not.  This naturally leads to a problem of testing multiple null hypotheses simultaneously.  If one were to test each of these null hypotheses in the usual way (i.e., ensuring that the probability of Type I error is controlled adequately for each null hypothesis separately), then the probability of {\it some} false rejection across all of the null hypotheses may be quite high.  A conventional solution to this problem is to require control of the familywise erorr rate -- the probability of {\it any} false rejection across all of the null hypotheses under consideration.  Methods for this problem for experiments in which treatment status is assigned in an i.i.d.\ fashion across units are developed in \cite{list2019multiple, list2023multiple}; see also \cite{lee2014multiple}.  A very complicated treatment assignment scheme that arises in the context of a well known experiment in the early childhood education literature is treated in \cite{heckman2024dealing}.  These results build upon general results in the multiple testing literature described in \cite{romano2005exact,romanoBalancedControlGeneralized2010}.  In some cases, especially when the number of null hypotheses under consideration is very large, it may be desirable to consider error rates that penalize false rejections less severely, such as the $k$ familywise error rate (defined as the probability of $k$ or more false rejections), the (tail probability of) the false discovery proportion (defined as the fraction of total rejections that are false), or the false discovery rate (defined as the expected value of the false discovery proportion).  Some relevant results for the control of such error rates are described in \cite{romanoBalancedControlGeneralized2010}; see also \cite{romano2007control} and \cite{romano2008formalized}.

\subsection{Imperfect Compliance and Attrition}
Even the most well designed experiments encounter challenges that can complicate subsequent analyses. Two common issues that arise in practice are imperfect compliance and attrition. 

Imperfect compliance arises when units assigned to the treatment group end up not taking up treatment, and/or units assigned to the control group manage to obtain the treatment. Of course, if researchers are simply interested in the effect of the \emph{assignment} to treatment as opposed to the true \emph{receipt} of the treatment (i.e., the intention-to-treat estimand), then everything we have discussed thus far applies directly. However, if the decision to comply with the treatment is not exogenous but is instead determined by the unobserved characteristics of the units, then the experiment no longer point identifies the causal effect of the true receipt of treatment on the outcome (i.e., the average treatment effect). Most recent work on the analysis of randomized experiments with imperfect compliance has adopted the framework of \cite{imbens1994identification}, where treatment assignment is used an an \emph{instrument} for the receipt of the treatment and the primary focus is on the so-called ``local" average treatment effect (i.e., the average treatment effect for those units who comply with the treatment assignment): see, in particular, \cite{ansel2018ols}, \cite{bugni2023inference}, \cite{ren2023model}, and \cite{bai2023inference-2}. Alternatively, a similar instrumental variables strategy could be used to \emph{partially} identify the average treatment effect: see \cite{manski1990nonparametric}, \cite{balke1997bounds}, \cite{bhattacharya2008treatment}, \cite{bhattacharya2012treatment}, \cite{machado2019instrumental}, \cite{bugni2024compliance}.

Attrition arises when outcomes are not observed for some subset of the experimental units. This situation could arise, for instance, if researchers lose track of subjects in the experiment. As with imperfect compliance, if the decision for the unit to drop out of the experiment is not exogenous, but is instead determined by their unobserved characteristics, then the experiment no longer point identifies the average treatment effect. Standard resolutions to this issue include modelling the selection process \citep[as in][]{heckman1979sample} or partial identification \citep[as in][]{horowitz2000nonparametric}. A textbook treatment on the analysis of randomized experiments with attrition is given in \cite{gerber2012}, and \cite{dinardo2006constructive} review different methods for dealing with attrition within the context of the Moving to Opportunity (MTO) experiment. \cite{ghanem2023testing} develop a test for attrition bias in randomized experiments. \cite{fukumoto2022nonignorable} and \cite{bai2024revisiting} study the problem of attrition in the setting of matched pair experiments, and revisit common recommendations about whether or not to drop pairs with an attrited unit.
\subsection{Network Experiments and Experiments with Interference}
Other than in Section \ref{sec:clust}, we have so far maintained that each individual's outcome depends on only their own treatment. A rapidly growing literature considers the analysis of experiments in the presence of \emph{interference}, i.e., that the outcome of a given individual in the experiment may be affected by the treatment statuses of others \citep[see][for an early discussion]{halloran1995causal}. The simplest example of such a setting is what is called \emph{partial} interference, where units are grouped into a collection of disjoint clusters, and interference is possible between individuals within the same cluster; Section \ref{sec:clust} discussed a special case of such a setting where every unit in the cluster receives the same treatment. \cite{hudgens2008toward}, \cite{basse2018analyzing}, \cite{imai2021causal}, \cite{vazquez2023identification}, \cite{leung2023design}, and \cite{liu2023inference} study \emph{two-stage experiments} in settings with partial interference, where first clusters are randomly assigned to different treatment \emph{proportions}, and then individuals within the clusters are assigned to treatment with (marginal) probability according to their cluster's assigned proportion. 


More generally, we could consider settings with complex patterns of interference, for instance, if individuals interact on a large network. \cite{manski2013identification} and \cite{aronow2017estimating} develop the concept of \emph{effective treatments} or \emph{exposure mappings}, which summarize how a given unit's outcome is affected by the treatments of other units. For example, in a network context, the exposure mapping may dictate that only the treatments of a unit's direct links affect their outcome. The exposure mapping formulation of interference has had a major influence on the subsequent literature: examples include \cite{leung2020treatment}, \cite{viviano2020experimental}, \cite{forastiere2021identification}, \cite{auerbach2021local}, \cite{munro2021treatment}, \cite{li2022random}, \cite{leung2022causal}, \cite{gao2023causal}, \cite{park2023assumption}, \cite{viviano2023causal}, \cite{savje2024causal}. Some recent papers that study general forms of interference without employing the formalism of exposure mappings are \cite{wager2021experimenting}, \cite{savje2021average}, \cite{hu2022average}, and \cite{faridani2024linear}.

\subsection{Randomization Inference}
An extremely important topic in the analysis of randomized experiments which we did not cover in this article is the idea of \emph{randomization inference}. Mechanically, randomization inference generates the null distribution of the test statistic by repeatedly re-assigning treatments to the experimental sample and re-computing the resulting test statistic. If the true value of the test-statistic is too ``large" relative to this null distribution then the null hypothesis is rejected. The primary strength of these types of tests is that, for appropriate null hypotheses, they can be shown to be \emph{finite sample} valid.

One such type of test, going back to the work of \cite{hoeffding1952large}, is where the data satisfies some form of \emph{group invariance} under the null hypothesis, with respect to a group of transformations of the data. For example, consider a completely randomized experiment where we wish to test the null hypothesis 
\[H_0: Y_i(1) \overset{d}{=} Y_i(0)~.\]
In this case, a valid group of transformations is given by the transformations which permute the treatment assignments of the individuals. \cite{lehmann2022testing} provide a comprehensive textbook introduction to these types of tests in very general settings (even outside the context of randomized experiments). \cite{romanoshaikhritzwoller2024} in this issue provide a survey of recent advances. 

A similar but distinct concept, going back to the work of \cite{fisher1925theory} and typically employed in the design-based paradigm, is where the null hypothesis is ``sharp" in the sense that the test statistic under any counterfactual treatment assignment can be imputed from the data. For example, consider a completely randomized experiment conducted on a finite population of $n$ individuals, then the canonical ``sharp" null is given by
\[H_0: Y_i(1) = Y_i(0)~ \text{  for all $1 \le i \le n $}~.\]
In this case, as we permute the treatment assignments of the individuals, we can perfectly impute the value of the test statistic under the null hypothesis. \cite{imbens2015causal} provide a comprehensive textbook introduction to these ``Fisher-style" tests. Recent work (particularly in settings with interference) includes \cite{athey2018exact}, \cite{basse2019randomization}, \cite{basse2024randomization}. 


It is important to emphasize that, in both cases, finite-sample validity is only guaranteed for \emph{very specific} choices of the null hypothesis. In particular, in either case considered above, if we instead consider the null hypothesis that the average treatment effect is equal to zero, then we would have neither the required  group invariance property nor a ``sharp" null which would guarantee finite-sample validity. However, a recent series of papers establishes that the randomization test can be \emph{asymptotically} valid for these ``weak" nulls, while retaining its finite-sample validity for the ``sharp" null, as long as the test-statistic is constructed appropriately. See \cite{chung2013exact}, \cite{bugni2018inference}, \cite{bai2022inference}, and \cite{bai2023inference-1}, for examples in a super-population context, and \cite{zhao2021covariate-adjusted}, \cite{wu2021randomization} for examples in a design-based context.


\subsection{Policy Learning}
Our discussion has focused on the problem of estimation and inference of treatment effect parameters. A recent literature, popularized in econometrics by \cite{manski2004statistical}, seeks to instead use the data in order to directly inform the allocation of treatments over the population as a function of the observable characteristics in order to maximize welfare. Formally, the problem is framed as a statistical decision problem in the framework of \cite{wald1949statistical} \citep[see][for a comprehensive introduction]{hirano2020asymptotic}. Extensive work on this topic, which is now often referred to as ``policy learning", exists at the intersection of econometrics, statistics, and computer science; important contributions to this literature which also provide comprehensive overviews are \cite{kitagawa2018should} and \cite{athey2021policy}. Relevant work in econometrics on this topic (primarily focused on settings where data are obtained from a randomized experiment) includes \cite{dehejia2005program}, \cite{stoye2009minimax}, \cite{hirano2009asymptotics}, \cite{bhattacharya2012inferring}, \cite{viviano2019policy}, \cite{ananth2020optimal}, \cite{azevedo2020b}, \cite{kitagawa2021equality}, \cite{mbakop2021model}, \cite{sun2021empirical}, \cite{viviano2022policy}, \cite{kitagawa2023should}, \cite{higbee2023policy}, \cite{kock2023treatment}. 

\subsection{Response-Adaptive Designs and Bandit Experiments}
Until now, we have assumed that the experimental design being employed by the researcher does not use information from earlier waves of experimentation. In this section, we briefly comment on a rapidly growing literature which considers \emph{response-adaptive} experimental designs, which are designs that can adapt throughout the experiment as a result of data that have already accrued. Response-adaptive designs have a long history in the analysis of clinical trials; \cite{hu2006theory} and \cite{rosenberger2015randomization} provide textbook introductions.

Some recent work on response-adaptive designs in econometrics and statistics studies procedures to construct feasible analogs of the Neyman allocation, in order to efficiently estimate treatment effect parameters; examples include \cite{hahn2011adaptive}, \cite{tabord-meehan2023stratification}, \cite{blackwell2022batch}, \cite{li2023double}, \cite{wei2024fair}. \cite{cai2022performance} demonstrate that some of these procedures may have poor finite-sample properties when the data used to estimate the optimal treatment assignment proportions is not sufficiently large.

Much of the recent work on adaptive designs is related to bandit problems and/or best arm identification. These are essentially policy learning problems (in the sense defined in Section 7.6) where the researcher wishes to choose a policy to maximize welfare, either for the participants in the experiment itself or for the broader population of interest. \cite{bubeck2012regret} and \cite{lattimore2020bandit} provide comprehensive textbook introductions. Some recent work on this topic includes \cite{russo2016simple}, \cite{russo2016information}, \cite{agrawal2017near}, \cite{kasy2021adaptive}, \cite{adusumilli2021risk}, \cite{lieber2022estimating}, \cite{kuang2023weak}, \cite{kato2024contextual}, \cite{higbee2024experimental}. The analysis of treatment effect parameters in these contexts can be particularly challenging since the experiment was not necessarily designed to facilitate inference. Recent work on inference in adaptive experiments includes \cite{bibaut2021post}, \cite{hadad2021confidence}, \cite{zhang2021statistical}, \cite{adusumilli2023optimal}, \cite{chen2023optimal}, \cite{hirano2023asymptotic}.

\clearpage
\appendix

\section{Additional Details}
\subsection{A General Comparison of Super-population and Finite-population Variances}\label{sec:variance_heuristic}
In this section we formalize a claim made in Section \ref{sec:finpop_intro} about the general relationship between the finite population and super-population variance of $\hat{\Delta}_n$.

When presenting the regularity conditions maintained in a finite-population analysis, researchers will often use the motivation that their conditions hold with probability one when the observations are in fact i.i.d.\ draws from a super-population \citep[see for instance the discussion following Theorem 5 in][]{li2017general}. With this in mind, finite-population results about the limiting distribution of $\hat{\Delta}_n$ can often be conceptualized as first imagining a collection of i.i.d.\ draws $W^{(N)} = \{(Y_i(1), Y_i(0), X_i): 1 \leq i \leq N\}$ from some distribution $P$, sampling a subset of $n$ units from $W^{(N)}$, and then deriving the limiting distribution of $\hat{\Delta}_n$ conditional on $W^{(N)}$:
\begin{equation} \label{eq:conditional-conv}
\sqrt n(\hat \Delta_n - \Delta_N) | W^{(N)} \stackrel{d}{\to} N(0, \sigma_{1, \lambda}^2(W^{(\infty)}))~,
\end{equation}
where $W^{(\infty)} = \{W_i: i \geq 1\}$ and $n / N \to \lambda \in [0, 1]$. Formally, this conditional convergence can be defined as
\[ \sup_{t \in \mathbf R} |P \{\sqrt n(\hat \Delta_n - \Delta_N) \leq t | W^{(N)}\} - \Phi(t / \sigma_{1, \lambda}(W^{\infty}))| \stackrel{P}{\to} 0~.\]
We now relate \eqref{eq:conditional-conv} to the unconditional limiting distribution of $\hat{\Delta}_n$ is a super-population analysis. First, we note that while in principle $\sigma_{1, \lambda}(W^{(\infty)})$ could vary with different realizations of the sequence of finite populations $W^{(\infty)}$, if $W^{(\infty)}$ in fact arise from a super-population, then $\sigma_1(W^{(\infty)})$ will often be the same for (almost all) realizations of $W^{(\infty)}$. As an example, if we consider complete randomization as defined in Section \ref{sec:setup} and if $W^{(\infty)}$ is drawn from a super-population $P$, then it follows from the strong law of large numbers that for almost every sequence $W^{(\infty)}$, as $n \to \infty$,
\begin{equation} \label{eq:fp-var-limit}
   n \var[\hat \Delta_n | W^{(N)}] \to \frac{\var_P[Y_i(1)]}{\pi} + \frac{\var_P[Y_i(0)]}{1 - \pi} - \lambda \var_P[Y_i(1) - Y_i(0)]~,
\end{equation}
where $n / N \to \lambda \in [0, 1]$ (we illustrate a similar property for stratified block randomization at the end of this section). Note that it follows immediately by the central limit theorem that $\sqrt n(\Delta_N - \Delta) \stackrel{d}{\to} N(0, \lambda \sigma_2^2)$, where $\sigma_2^2 = \var_P[Y_i(1) - Y_i(0)]$. Because $\Delta_N$ is a function of $W^{(\infty)}$, if \eqref{eq:conditional-conv} holds with $\sigma_{1, \lambda}^2(W^{(\infty)}) \equiv \sigma_{1, \lambda}^2$, then it follows from Lemma S.1.2 in \cite{bai2022inference} that the distributions of $\sqrt n(\hat \Delta_n - \Delta_N) | W^{(N)}$ and $\sqrt n(\Delta_N - \Delta)$ are ``asymptotically independent'' and moreover that
\begin{equation} \label{eq:uncond-conv}
 \sqrt n (\hat \Delta_n - \Delta) \stackrel{d}{\to} N(0, \sigma_{1, \lambda}^2 + \lambda \sigma_2^2)~. 
 \end{equation}
 In other words, the super-population variance given in \eqref{eq:uncond-conv} and the finite-population variance given in \eqref{eq:conditional-conv} differ by $\lambda \var_P[Y_i(1) - Y_i(0)]$ in the limit, which is the fraction of people sampled from the finite population times the variance of the individual level treatment effects. When $n / N \to \lambda = 0$, so that we sample a negligible fraction of the finite population in the limit, the difference is zero. Indeed, we can see this for complete randomization by comparing \eqref{eq:completerand-super} and \eqref{eq:fp-var-limit}. It thus follows immediately that constructing a consistent estimator of $\sigma_{1, \lambda}^2 + \lambda \sigma^2_2$ delivers a conservative variance estimator of $\sigma_{1, \lambda}^2$ (in fact, $\sigma_{1, \lambda} + \lambda \sigma_2^2$  doesn't vary across $\lambda$, as can be seen for example in complete randomization.)

As another example, consider stratified block randomization with discrete $x \in \{1, 2, \ldots, \mathcal{X}\}$ as defined in Section \ref{sec:sbr}. Here we suppose $n = N$ so that $\lambda = 1$. For $d \in \{0, 1\}$ and $x \in \{1, 2, \ldots, \mathcal{X}\}$, define
\begin{align*}
    n_d(x) & = \sum_{1 \leq i \leq n} I \{D_i = d, X_i = x\} \\
    n(x) & = n_1(x) + n_0(x) \\
    \bar Y_n(d, x) & = \frac{1}{n_d(x)} \sum_{1 \leq i \leq n} Y_i(d) I \{D_i = d, X_i = x\} \\
    \Delta_n(x) & = \bar Y_n(1, x) - \bar Y_n(0, x)~.
\end{align*}
The finite population variance of the fully saturated estimator $\hat \Delta_n^{\rm sat}$ is
\[ \var[\hat \Delta_n^{\rm sat} | W^{(N)}] = \sum_{x} \frac{n(x)}{n} \left ( \frac{S_1^2(x)}{n_1(x)} + \frac{S_0^2(x)}{n_0(x)} - \frac{S_\Delta^2(x)}{n(x)} \right )~, \]
where $S_d^2(x)$ and $S_\Delta^2(x)$ are the within-stratum counterpart of $S_d^2$ and $S_\Delta^2$ in \eqref{eq:fp_var}:
\begin{align*}
    S_d^2(x) & = \frac{1}{n_d(x) - 1} \sum_{1 \leq i \leq n} (Y_i(d) - \bar Y_n(d, x))^2 I \{D_i = d, X_i = x\} \\
    S_\Delta^2(x) & = \frac{1}{n(x) - 1} \sum_{1 \leq i \leq n} (Y_i(1) - Y_i(0) - \Delta_n(x))^2 I \{X_i = x\}~.
\end{align*}
If the finite populations are realizations from a super-population $P$, then it can be shown that with probability one,
\[ n \var[\hat \Delta_n^{\rm sat} | W^{(N)}] \to \sum_{x} p(x) \left ( \frac{\var[Y_i(1) | X_i = x]}{\pi(x)} + \frac{\var[Y_i(0) | X_i = x]}{1 - \pi(x)} - \var[Y_i(1) - Y_i(0) | X_i = x] \right )~, \]
where $p(x) = P \{X_i = x\}$.

On the other hand, the super-population variance satisfies
\begin{align*}
    n \var[\hat \Delta_n^{\rm sat}] & \to \sum_{x} p(x) \bigg ( \frac{\var[Y_i(1) | X_i = x]}{\pi(x)} + \frac{\var[Y_i(0) | X_i = x]}{1 - \pi(x)} \\
    & \hspace{5cm} + (E[Y_i(1) - Y_i(0) | X_i = x] - E[Y_i(1) - Y_i(0)])^2 \bigg )~.
\end{align*}
Therefore, once again we have
\[ n \var[\hat \Delta_n^{\rm sat}] - n \var[\hat \Delta_n^{\rm sat} | W^{(N)}] \to \var[Y_i(1) - Y_i(0)]~. \]

\section{Proofs of claims in main text}
\subsection{Derivations of $E[\hat{\Delta}_n]$ and $\var[\hat{\Delta}_n]$ under complete randomization in the finite population framework}\label{sec:fin_bias_var}
Define the set $\mathcal{C}_{n,N}$ to index the random subset of $n$ observations sampled without replacement from the population of size $N$. Consider the following expansion:
\[\hat{\Delta}_n = A_n - B_n~,\]
where
\[A_n = \frac{1}{n_1}\sum_{1 \le i \le n}\left(Y_i(1) + Y_i(0)\frac{n_1}{n_0}\right)D_i\]
\[B_n = \frac{1}{n_0}\sum_{1 \le i \le n}Y_i(0)~.\]
In this re-writing, we have partitioned $\hat{\Delta}_n$ into two components $A_n$ and $B_n$. Conditional on $\mathcal{C}_{n,N}$, and given our definition of $D^{(n)}$, the first component $A_n$ can be interpreted as the sample average when sampling $n_1$ units without replacement from the finite population of observations  $\{(Y_i(1) + Y_i(0)\frac{n_1}{n_0}): 1 \le i \le n\}$, whose population average is given by
\[\frac{1}{n}\sum_{1 \le i \le n}\left(Y_i(1) + Y_i(0)\frac{n_1}{n_0}\right)~.\]
When viewed from this perspective, the treatment indicators $D_i$ are now \emph{sampling} indicators which determine which of the $n_1$ units are sampled from our population of $n$ units. Using standard results from the literature on survey sampling \citep[see, for instance, Theorem 2.1 in ][]{cochran1977sampling}, the sample average is unbiased for the population average:
\[E[A_n|\mathcal{C}_{n,N}] = \frac{1}{n}\sum_{1 \le i \le n}\left(Y_i(1) + Y_i(0)\frac{n_1}{n_0}\right)~,\]
where we emphasize that the expectation is with respect to the sampling indicators $D^{(n)}$, which are indeed the only source of randomness once we condition on $\mathcal{C}_{n,N}$. 
Combining this result with our decomposition for $\hat{\Delta}_n$, we obtain that
\begin{align*}
E[\hat{\Delta}_n|\mathcal{C}_{n,N}] &= E[A_n|\mathcal{C}_{n,N}] - E[B_n|\mathcal{C}_{n,N}] \\
&= \frac{1}{n}\sum_{1 \le i \le n}\left(Y_i(1) + Y_i(0)\frac{n_1}{n_0}\right) - B_n \\
& = \frac{1}{n}\sum_{1 \le i \le n}\left(Y_i(1) - Y_i(0)\right)~,
\end{align*}
where the second equality used our above derivation of $E[A_n|\mathcal{C}_{n,N}]$ and the fact that $B_n$ is non-random conditional on $\mathcal{C}_{n,N}$, and the third equality follows from some additional algebra. Then by the law of iterated expectations:
\[E[\hat{\Delta}_n] = E[E[\hat{\Delta}_n|\mathcal{C}_{n,N}]] = E\left[\frac{1}{n}\sum_{1 \le i \le n}\left(Y_i(1) - Y_i(0)\right)\right] = \Delta^{\rm fp}_N~,\]
where the last equality again follows from Theorem 2.1 in \cite{cochran1977sampling}.
Using the same decomposition, we obtain that
\begin{align*}
\var[\hat{\Delta}_n|\mathcal{C}_{n,N}] &= \var[A_n|\mathcal{C}_{n,N}]\\
&=\var\left[\frac{1}{n_1}\sum_{1 \le i \le n} \left(Y_i(1) + Y_i(0) \frac{n_1}{n_0}\right)D_i \right] \\
&= \left(\frac{1}{n_1} - \frac{1}{n} \right)
						\frac{1}{n-1} \sum_{1 \le i \le n} \left[
							 \left(Y_i(1) + Y_i(0) \frac{n_1}{n_0}\right)
							 - \frac{1}{n} \sum_{1 \le i \le n}  \left(Y_i(1) + Y_i(0) \frac{n_1}{n_0}\right)
							 \right]^2 \\
        & = \frac{\zeta_1^2}{n_1} + \frac{\zeta_0^2}{n_0} - \frac{1}{n(n-1)} \sum_{1 \le i \le n} \left[(Y_i(1) -\bar{Y}_n(1))^2 + (Y_i(0) - \bar{Y}_n(0))^2 - 2 (Y_i(1) - \bar{Y}_n(1))(Y_i(0) - \bar{Y}_n(0)) \right] \\
       &= \frac{\zeta_1^2}{n_1} + \frac{\zeta_0^2}{n_0} - \frac{\zeta_\Delta^2}{n}~,
\end{align*}
with
\begin{align*}
\zeta^2_d &= \frac{1}{n-1}\sum_{1 \le i \le n}\left(Y_i(d) -\bar{Y}_n(d)\right)^2 \\
\zeta^2_\Delta &= \frac{1}{n-1}\sum_{1 \le i \le n}\left(Y_i(1) - Y_i(0) - \left(\bar{Y}_n(1) - \bar{Y}_n(0)\right)\right)^2~,
\end{align*}
where the first equality follows since $B_n$ is non-random conditional on $\mathcal{C}_{n,N}$, the third equality follows from Theorem 2.2 in \cite{cochran1977sampling} and the last two equalities from additional algebra. We thus obtain from the law of total variance that
\begin{align*}
\var[\hat{\Delta}_n] &= E[\var[\hat{\Delta}_n|\mathcal{C}_{n,N}]] + \var[E[\hat{\Delta}_n|\mathcal{C}_{n,N}]]\\
&=E\left[\frac{\zeta_1^2}{n_1} + \frac{\zeta_0^2}{n_0} - \frac{\zeta_\Delta^2}{n}\right] + \var\left[\frac{1}{n}\sum_{1 \le i \le n}\left(Y_i(1) - Y_i(0)\right)\right]~. 
\end{align*}
Repeated applications of Theorems 2.1 and 2.2 in \cite{cochran1977sampling} along with additional algebra reveals that
\[E\left[\zeta^2_d\right] = \frac{n}{n-1}\left(\left(\frac{N-1}{N}\right)S^2_d - \left(\frac{1}{n} - \frac{1}{N}\right)S^2_d\right) = S^2_d~,\]
and similarly
\[E\left[\zeta^2_\Delta\right] = S^2_\Delta~.\]
By another application of Theorem 2.2 in \cite{cochran1977sampling},
\[\var\left[\frac{1}{n}\sum_{1 \le i \le n}\left(Y_i(1) - Y_i(0)\right)\right] = \left(\frac{1}{n} - \frac{1}{N}\right)S^2_{\Delta}~.\]
Putting this all together, we obtain
\begin{align*}
\var[\hat{\Delta}_n] &= \frac{S^2_1}{n_1} + \frac{S^2_1}{n_0} - \frac{S^2_\Delta}{n} + \left(\frac{1}{n} - \frac{1}{N}\right)S^2_\Delta \\
&= \frac{S^2_1}{n_1} + \frac{S^2_1}{n_0} - \frac{S^2_\Delta}{N}~,
\end{align*}
as desired. \qed

\subsection{Derivations of \eqref{eq:dim_unbiased} and \eqref{eq:super_var}}
For the first claim,
\begin{align*}
E[\hat\Delta_n] &= E\left[E\left[\hat\Delta_n|D^{(n)}\right]\right] \\
         &= E\left[\frac{1}{n_1}\sum_{1 \le i \le n}E[Y_i(1)|D^{(n)}]D_i - \frac{1}{n_0}\sum_{1 \le i \le n}E[Y_i(0)|D^{(n)}](1 - D_i)\right] \quad \\
         &= E\left[E[Y_i(1)]\left(\frac{1}{n_1}\sum_{1 \le i \le n}D_i\right) - E[Y_i(0)]\left(\frac{1}{n_0}\sum_{1 \le i \le n}(1 - D_i)\right)\right] \quad \\
         & = E[Y_i(1) - Y_i(0)] = \Delta~,
\end{align*}
where the first equality follows from the law of iterated expectations, the second from properties of conditional expectations and the definition of the observed outcome $Y_i$, the third from the exogeneity of treatment assignment and the fact that the sample is i.i.d., and the final equality by the definition of $n_1$ and $n_0$.

\noindent For the second claim,
\begin{align*}
\text{Var}[\hat\Delta_n] &= E\left[\text{Var}\left[\hat\Delta_n|D^{(n)}\right]\right] + \text{Var}\left[E\left[\hat{\Delta}_n|D^{(n)}]\right]\right] \quad\\
& = E\left[\text{Var}\left[\hat\Delta_n|D^{(n)}\right]\right] + \text{Var}[\Delta] \quad \\
& = E\left[\frac{1}{n_1^2}\sum_{1 \le i \le n}\text{Var}[Y_i(1)|D^{(n)}]D_i + \frac{1}{n_0^2}\sum_{1 \le i \le n}\text{Var}[Y_i(0)|D^{(n)}](1 - D_i)\right] \quad \\
& = E\left[\text{Var}[Y_i(1)]\frac{1}{n_1^2}\sum_{1 \le i \le n}D_i + \text{Var}[Y_i(0)]\frac{1}{n_0^2}\sum_{1 \le i \le n}(1 - D_i)\right] \quad \\
& = \frac{\text{Var}[Y_i(1)]}{n_1} + \frac{\text{Var}[Y_i(0)]}{n_0}~, 
\end{align*}
where the first equality follows from the law of total variance, the second from the derivation of the expectation, the third from the properties of conditional variances and the fact that $D_i$ is binary, the fourth by the exogeneity of treatment assignment and the fact that the sample is i.i.d., and the final equality by the definition of $n_1$ and $n_0$ under complete randomization.
\subsection{Derivation of \eqref{eq:EVarDelta}}
We have
\begin{align*}
   & E[\var[\hat \Delta_n | X^{(n)}, D^{(n)}] | X^{(n)}] \\
   & = E \bigg [ \frac{4}{n^2} \sum_{1 \leq i \leq n} (D_i \var[Y_i(1) | X_i] + (1 - D_i) \var[Y_i(0) | X_i]) \bigg | X^{(n)} \bigg ] \\
   & = \frac{2}{n^2} \sum_{1 \leq i \leq n} (\var[Y_i(1) | X_i] + \var[Y_i(0) | X_i])~,
\end{align*}
where in the first equality we use the fact that the potential outcomes are independent across units conditional on $X^{(n)}$ and $D^{(n)}$. \qed

\clearpage
\bibliography{all}
\end{document}